\documentclass[preprint]{aastex}

\usepackage{natbib}
\usepackage{amsmath}
\usepackage{graphicx}

\shorttitle{The Evolved MS Channel: CSS1111}
\shortauthors{Kennedy et al.}

\begin{document}


\title{The Evolved Main-Sequence Channel: \\HST and LBT observations of CSS120422:111127+571239}


\author{M. Kennedy\altaffilmark{1}$^{,}$\altaffilmark{2}, P. Garnavich\altaffilmark{2}, P. Callanan\altaffilmark{1}, P. Szkody\altaffilmark{3}, C. Littlefield\altaffilmark{2}$^{,}$\altaffilmark{4}, R. Pogge\altaffilmark{5}$^{,}$\altaffilmark{6}}

\altaffiltext{1}{Department of Physics, University College Cork, Cork, Ireland}
\altaffiltext{2}{Department of Physics, University of Notre Dame, Notre Dame, IN 46556}
\altaffiltext{3}{Department of Astronomy, University of Washington, Seattle, WA}
\altaffiltext{4}{Wesleyan University, Astronomy Department, Van Vleck Observatory, 96 Foss Hill Dr. Middletown, CT 06459}
\altaffiltext{5}{Department of Astronomy, The Ohio State University, 140 W. 18th Avenue, Columbus, OH 43202, USA}
\altaffiltext{6}{Center for Cosmology and AstroParticle Physics, The Ohio State University, 191 West Woodruff Avenue, Columbus, OH 43210, USA}

\begin{abstract}
The ``evolved main-sequence'' channel is thought to contribute significantly to the population of AM CVn type systems in the Galaxy, and also to the number of cataclysmic variables detected below the period minimum for hydrogen rich systems. CSS120422:J111127+571239 was discovered by the Catalina Sky Survey in April 2012. Its period was found to be 56 minutes, well below the minimum, and the optical spectrum is clearly depleted in hydrogen relative to helium, but still has two orders of magnitude more hydrogen than AM CVn stars. Doppler tomography of the H$\alpha$ line hinted at a spiral structure existing in the disk. Here we present spectroscopy of CSS120422:J111127+571239 using the COS FUV instrument on the Hubble Space Telescope and using the MODS spectrograph on the Large Binocular Telescope. The UV spectrum shows SiIV, NV and HeII, but no detectable CIV. The anomalous nitrogen/carbon ratio is seen in a small number of other CVs and confirms a unique binary evolution. We also present and compare the optical spectrum of V418 Ser and advocate that it is also an evolved main-sequence system.
\end{abstract}


\section{Introduction}

Cataclysmic Variables (CVs) are close binary systems composed of a white dwarf primary star which is typically accreting material via Roche Lobe (RL) overflow from a main sequence companion. The period distribution of CVs shows very few systems with a period below $P\approx78$ min, which is the period minimum for systems which have a main sequence (MS) companion \citep{kolb2002}. In short period binaries, angular momentum loss is caused by the emission of gravitational radiation, which shortens the orbital period of the system. However, when the thermal time scale of the donor star becomes comparable to the angular momentum loss time scale, the period can increase, resulting in a period minimum (\citealt{Paczynski1981}; \citealt{Paczynski1981b}). There are close binary systems which have periods below this minimum, suggesting that there are ways to avoid the minimum.

There are currently 3 different channels to create systems which have a period below the period minimum: 1) The system is double degenerate. That is, both objects are white dwarfs (WD), with the more massive WD accreting material from the less massive WD (\citealt{Tutukov79}; \citealt{Nather81}). 2) The companion is a helium burning star, which means the system has a much shorter minimum period (\citealt{Savonije86}; \citealt{Iben1991}). 3) The companion evolved off of the main sequence just as mass transfer began. This last channel is known as the evolved main sequence (EMS) channel \citep{Podsiadlowski2003}, which was initially proposed by \cite{Tutukov1985} and \cite{Nelson1986}.

Early work on generating populations of CVs with periods below the minimum found that the first 2 of these channels were important, and that the EMS channel did not contribute to the overall population \citep{Nelemans2001}. Since then, more in-depth modelling has found that the choice of efficiency in ejecting material during the common-envelope stage has a dramatic effect on which channels are important, and that for inefficient common-envelope ejection, the EMS channel contributes significantly to systems with short periods \citep{Podsiadlowski2003}.

The mechanism for forming an ultracompact system by the EMS channel requires a very particular set of starting conditions. Firstly, the mass of the companion must initially be greater than that of the white dwarf primary i.e. $q>1$. If this condition is satisfied, then after the common envelope stage, the system can undergo thermal timescale mass transfer (TTMT), as the Roche Lobe radius ($R_{L}$) shrinks faster or at the same rate as its thermal equilibrium radius ($R_{th}$) \citep{Tutukov79}. Once enough mass has been lost by the secondary and $q$ has become small enough, $R_{L}$ shrinks slower than $R_{th}$, and the system detaches when $R_{th}$ drops below $R_{L}$. 

Secondly, the companion star must have just finished hydrogen burning when the system becomes semi-detached directly after the common envelope stage. If this is true, then as the system evolves, the star becomes fully convective after it has lost more mass than if it had been a main sequence star. Thus magnetic braking becomes ineffective at smaller periods and the period of detachment (known as the period gap) is much shorter than in conventional CVs \citep{Podsiadlowski2003}. It follows that the system undergoes high rates of mass transfer for longer times than conventional CVs. The period minimum is also affected by having an evolved companion. The modelling done by \cite{Podsiadlowski2003} shows that if the secondary is evolved and has a helium core surronded by a hydrogen envelope, then the period minimum can drop down to $P_{min}\approx 11$ minutes for a helium core mass of $0.063\:M_{\odot}$.

Identification of a short period CV formed by the EMS channel should be easy. First, as opposed to double degenerate or helium star channels, EMS channel systems should have detectable hydrogen in their spectrum, even when the system is below the period minimum for main sequence companions. In most of the evolutionary tracks simulated by \cite{Podsiadlowski2003}, they found that the efficiency of magnetic braking, and by proxy, the evolutionary state of the companion, affects the value of the period minimum. For most of the tracks, when the period minimum is $>$ 11 min, the EMS systems still have detectable H after reaching the period minimum. However, for the most extreme case of magnetic braking efficiency, where the period minimum is $\approx$ 11 min,  the hydrogen is exhausted from the companion at the period minimum, and the EMS systems that evolve along this path become similar to the systems formed by the first 2 channels, which is supported by the abundance ratios calculated by \cite{Nelemans2010}. Secondly, these systems are predicted to have very high N/C ratios, as the CNO processed material in the hydrogen shell, which is normally hidden in stars, should be at the surface due to the high rates of accretion that these systems undergo (\citealt{Marks1998}; \citealt{Podsiadlowski2003}; \citealt{Harrison05}; \citealt{Nelemans2010}). 

The first observed inverted N/C ratio system was BY Cam \citep{Bonnet1987}, and was thought to be due to anomalous abundances from either a nova outburst or an evolved main sequence star. Since then, more CV systems have displayed this anomaly in the UV (\citealt{Gansicke2003}; \citealt{Sanad2011}) and it has also been proposed that this anomaly can be detected in the IR \citep{Hamilton2011}. There is a bottleneck in the carbon-nitrogen cycle, which sees the majority of $^{12}$C close to the core of MS stars processed into $^{14}$N, resulting in the an inverted N/C ratio close to the core of MS stars. Typically, this layer is not observable. However, in the EMS channel, the outer hydrogen layer is stripped away during the period of TTMT. This exposes the inner hydrogen layer above the helium core which has been nitrogen enriched via this process, leading to the inverted N/C ratio (\citealt{Schenker2002}; \citealt{Gansicke2003}). Out of the sample observed by \cite{Gansicke2003}, only 2 objects had a period below the period minimum - EI Psc (RXJ2329.8+0628) with $P=64.176$ min \citep{Thorstensen2002} and CE 315 with $P=65.1$ min \citep{Ruiz2001}. CE 315 has been confirmed to be a double degenerate system, but EI Psc shows hydrogen emission lines alongside strong He lines, suggesting that EI Psc is a CV following the EMS channel.

CSS120422:J111127+571239, also known as SBSS 1108+574 and SDSS J111126.84+571238.6 and hereby referred to as CSS1111, is a CV system discovered by the Catalina Sky Survey when it went into outburst in April 2012. It was observed to have superhumps with an ultrashort period of 0.038449(6) days \citep{Kato2013}. \cite{Carter2013} reported a spectroscopic period of $55.3\pm0.8$ min using radial velocity measurements, and found evidence of a possible eclipse of the accretion disk based on observations of the HeII$_{\lambda4686}$ line. The orbital period was confirmed to be 55.36 minutes using photometry obtained in 2012 while the optical spectrum of the object showed hydrogen emission lines alongside the strong helium lines expected in an AM CVn system (\citealt{Carter2013}; \citealt{Littlefield2013}). Due to the high N/C ratio expected from EMS systems, the most important part of the spectrum in classifying CSS1111 is the UV, due to the presence of the NV$_{\lambda1238}$ and CIV$_{\lambda1550}$ lines. The Cosmic Origins Spectrograph on the Hubble Space Telescope is the perfect instrument for observing these far UV lines.

V418 Ser is a faint (m$>$20) cataclysmic variable that was discovered by the ROTSE-III telescope in 2004 after going into outburst and reaching a magnitude of 15.4 \citep{Rykoff2004}. The CRTS observed V418 Ser to be in outburst again in May 2014, and extensive photometry found a photometric period of 64.2(5) minutes, which has been attributed to the superhump period of the system(Littlefield, VSNET alert 17321; de Miguel, VSNET alert 17322). The orbital period should lie within a few $\%$ of the superhump period.

Here, we present the first UV data, as well as new optical spectroscopy taken after an extended period of quiescence, to investigate whether CSS1111 is indeed a product of the EMS channel. We compare our data to previous observations, search for possible eclipses in the UV continuum or emission lines and determine whether the spiral structure in the disk at the 2:1 resonance as found by \cite{Littlefield2013} was still present in 2014. We also present optical spectroscopy of V418 Ser, and confirm that it is also a product of the EMS channel. Finally, we discuss the relationship between EMS channel systems and Type Ia supernovae.

\section{Observations and Analysis}

Observations of CSS1111 were obtained using the Hubble Space Telescope (HST), the Large Binocular Telescope (LBT), the Vatican Advanced Technology Telescope (VATT) and Apache Point Observatory (APO) as detailed in the sections below. Table \ref{obs_det} shows a summary of the observations.

\begin{deluxetable}{cccccccc}
\tabletypesize{\small}
\tablecolumns{8}
\tablewidth{0pc}
	\tablecaption{Details of Observations of CSS1111 and V418 Ser\label{obs_det}}
	\tablehead{
		\tableline
		\colhead{Date}	&\colhead{Object}	&\colhead{Obs}	&\colhead{Filter}	&\colhead{Exp(s)}	&\colhead{UT Time}	&\colhead{Phase}
		}
		
		\startdata
		24 April 2014 &CSS1111 &LBT &MODS1 & 200 &05:52:49&0.21\\
		24 April 2014 &CSS1111 &LBT &MODS1 & 200 &05:57:52&0.30\\
		24 April 2014 &CSS1111 &LBT &MODS1 & 200 &06:02:55&0.39\\
		24 April 2014 &CSS1111 &LBT &MODS1 & 200 &06:13:04&0.57\\
		24 April 2014 &CSS1111 &LBT &MODS1 & 200 &06:18:06&0.66\\
		24 April 2014 &CSS1111 &LBT &MODS1 & 200 &06:23:09&0.75\\
		24 April 2014 &CSS1111 &LBT &MODS1 & 200 &06:28:12&0.85\\
		24 April 2014 &CSS1111 &LBT &MODS1 & 200 &06:33:14&0.94\\
		24 April 2014 &CSS1111 &LBT &MODS1 & 200 &06:38:17&0.03\\
		24 April 2014 &CSS1111 &LBT &MODS1 & 200 &06:43:23&0.12\\
		24 April 2014 &CSS1111 &LBT &MODS1 & 200 &06:48:25&0.21\\
		24 April 2014 &CSS1111 &LBT &MODS1 & 200 &06:53:28&0.30\\
		24 April 2014 &CSS1111 &LBT &MODS1 & 200 &06:58:30&0.39\\
		24 April 2014 &CSS1111 &LBT &MODS1 & 200 &07:03:32&0.48\\
		24 April 2014 &CSS1111 &LBT &MODS1 & 200 &07:08:35&0.57\\
		29 April 2014 &CSS1111 &VATT &sdss-g &265 Images& &\\
		01 May 2014 &CSS1111 &VATT &sdss-g &198 Images& &\\
		02 May 2014 &CSS1111 &VATT &sdss-g &322 Images& &\\
		16 May 2014 &CSS1111 &HST &G140L Standard & 2754 &06:30:41&0.09-0.92\\
		16 May 2014 &CSS1111 &HST &G140L Standard & 2000 &07:53:09&0.59-0.18\\
		16 May 2014 &CSS1111 &HST &G140L Standard & 1081 &08:28:33&0.22-0.54\\
		16 May 2014 &CSS1111 &HST &G140L Standard & 1000 &09:29:06&0.31-0.61\\
		16 May 2014 &CSS1111 &HST &G140L Standard & 2090 &09:47:41&0.64-0.27\\
		16 May 2014 &CSS1111 &HST &G140L Shifted & 3205 &11:05:03&0.04-0.00\\
		16 May 2014 &CSS1111 &HST &G140L Shifted & 2000 &12:40:58&0.77-0.38\\
		16 May 2014 &CSS1111 &HST &G140L Shifted & 1084 &13:16:22&0.41-0.74\\
		16 May 2014 &CSS1111 &HST &G140L Shifted & 1000 &14:16:52&0.51-0.81\\
		16 May 2014 &CSS1111 &HST &G140L Shifted & 2090 &14:35:27&0.84-0.47\\
		30 June 2014 &V418 Ser &LBT &MODS1 & 300 &05:08:49&\\
		30 June 2014 &V418 Ser &LBT &MODS1 & 300 &05:15:30&\\
		30 June 2014 &V418 Ser &LBT &MODS1 & 300 &05:22:12&\\
		30 June 2014 &V418 Ser &LBT &MODS1 & 300 &05:28:53&\\
		30 June 2014 &V418 Ser &LBT &MODS1 & 300 &05:35:35&\\
		\enddata
\end{deluxetable}

\subsection{HST Observations}
Observations of CSS1111 were taken during six HST orbits on 2014 May 16 using the Cosmic Origins Spectrograph (COS). A total of 10 spectra were obtained in TIME-TAG mode using the G140L grating. The G140L grating has a dispersion of $80$ m\AA$\:$ pixel$^{-1}$, with a resolving power of $R=1500-4000$ in the bandpass $950-2150$\AA$\:$. The grating was in the standard position for the first 3 orbits, covering the range 1130\AA$\:$  to 2000\AA$\:$. For the last 3 orbits, the grating was shifted such that the spectral range went from 900\AA$\:$ to 1200\AA$\:$ and 1275\AA$\:$ to 2000\AA$\:$. The phase coverage of each observation can be seen in Table 1, and is based on the ephemeris calculated later in this paper.

The spectra and TIME-TAG data were analysed using a combination of PyRAF routines from the \textsc{STSDAS} package \textsc{HSTCOS} (version 3.17) and basic \textsc{PYTHON} procedures. Light curves were made for various parts of the UV spectrum to look for an eclipse of the accretion disk, which had been possibly seen in the HeII$_{\lambda4686}$ line during outburst \citep{Carter2013}. Since the count rate for the system changed when the grating was shifted, the light curves were normalized, and the variations in the light curve are expressed as a fraction of the median number of counts.

If CSS1111 had been in outburst (which happens approximately once per year, see Section \ref{EvolvingDisk} for more information) during the HST observations, the flux would have been too bright for the COS FUV. As such, the American Association of Variable Star Observers (AAVSO) provided observations leading up to and during the HST observations to confirm that CSS1111 was in quiescence. 

\subsection{LBT Observations}
Fifteen optical spectra of CSS1111 were obtained using the MODS spectrograph \citep{PoggeMODS} on the Large Binocular Telescope (LBT) on 2014 April 23, 23 days before the HST observations. The MODS spectrograph has 2 gratings, one blue grating covering the wavelength range $3200-5800$\AA$\:$ with a dispersion of $0.5$ \AA$\:$ pixel$^{-1}$ and one red grating covering the wavelength range $5800-10000$\AA$\:$ with a dispersion of $0.8$ \AA$\:$ pixel$^{-1}$. Both gratings provide a resolution of $R\approx2000$. The 1.2-arcsec segmented long-slit was used to obtain spectra, each with an exposure time of 200s. The spectra were flat fielded, debiased and wavelength and flux calibrated using calibration data from the same run. The wavelength calibration is based on observations of Argon, Krypton and  Xenon lamps for the blue arm, Neon and Argon lamps for the red and standard LBT line lists. The data reduction tasks were carried out using PyRAF and IRAF \footnote{IRAF is distributed by the National Optical Astronomy Observatory, which is operated by the Association of Universities for Research in Astronomy (AURA) under cooperative agreement with the National Science Foundation}.

Archive observations of CSS1111 taken with the LBT in 2012, as detailed in \citet{Littlefield2013}, were also reduced and analysed, so that a direct comparison between the spectrum in 2014 and 2012 could be made.

Five optical spectra of V418 Ser were obtained using the MODS spectrograph on 2014 June 30, each with an exposure time of 300s. The spectra were reduced in the same manner as the CSS1111 spectra. 

\subsection{VATT Observations}
\label{VATT}
Photometry of CSS1111 was taken in the SDSS \textit{g} band using the 1.8m Vatican Advanced Technology Telescope (VATT) located at Mount Graham International Observatory on 3 separate nights - 2014 April 29, May 1 and May 2. The images were taken using the VATT4K CCD with a typical exposure time of 30s. A total of 785 images were taken over these 4 nights.

\subsection{APO Observations}
Nine optical spectra of CSS1111 were obtained on 2014 May 4 and 6, 12 and 10 days prior to the HST observations, using the Dual Imaging Spectrograph (DIS) on the 3.5m ARC telescope at Apache Point Observatory. The blue channel used the B1200 grating, which has a pixel scale of 0.62 \AA$/$pix, while the red channel used the R1200 grating, which has a pixel scale of 0.58 \AA$/$pix. While these spectra will not be discussed in detail as the S/N is poorer than the LBT spectra, they do confirm the quiescent state of the system and have similar properties to the LBT data.

\section{Results}

\subsection{Optical Spectrum}
\begin{figure}
\epsscale{1}
\plotone{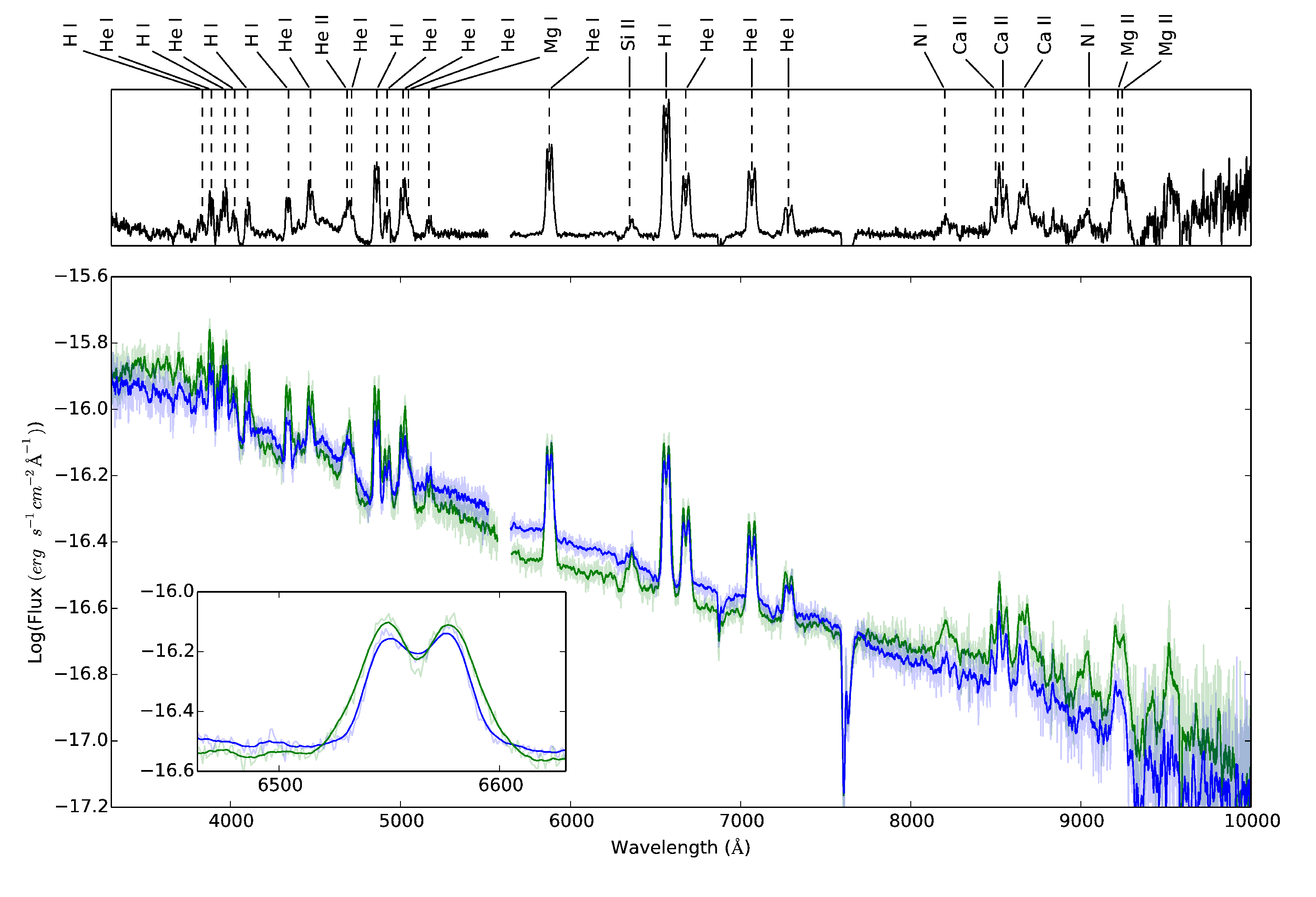}
\caption{Average spectrum of CSS1111 using the MODS1 spectrograph on the LBT. The blue line is the data taken in 2012, and the green line is from 2014. The inset compares the spectral differences between 2012 and 2014 for H$\alpha$, most notably the wider wings of the lines, suggesting changes in the accretion disk. The transparent green and blue lines represent the unsmoothed averaged spectrum. The top panel shows the 2014 and 2012 data averaged together and normalised, to make line identification easier.\label{average_spec}}
\end{figure}

The average optical spectrum taken with the LBT in both 2012 and 2014 can be seen in Figure \ref{average_spec}. The strongest emission lines have been marked, which include the Balmer series, HeI and HeII, SiII, NI and MgI and MgII. The NI and MgI and MgII lines were identified by comparison with the spectrum of the disk in GP Comae \citep{Marsh1991}. The lab wavelengths of the NI lines are $8200$\AA$\:$ and $9050$\AA$\:$ versus the observed wavelengths of $8206$\AA$\:$ and $9037$\AA$\:$. Like \cite{Marsh1991}, our identification of NI is not exact, but may be due to the large line width and complexity of the spectrum in these regions, making accurate measurements difficult. There have been significant changes in the spectrum since 2012. Every emission line has a higher flux and higher equivalent width (see Table \ref{opt_ew}). Also, as shown in the inset of Figure \ref{average_spec}, the emission features have become more prominently double peaked with wider emission wings. The peak-to-peak separation of the H$\alpha$ line is unchanged at 27\AA$\:$.

\begin{deluxetable}{rccccc}
\tabletypesize{\tiny}
\tablecolumns{6}
\tablewidth{0pc}
\rotate 
	\tablecaption{Equivalent width and fluxes of some of the most prominent emission lines in the optical spectrum for both CSS1111 and V418 Ser. \label{opt_ew}}
	\tablehead{
		\tableline
		\colhead{}			&\colhead{H$\alpha$} 			& \colhead{H$\beta$} 				& \colhead{HeI (5876 \AA)} 				& \colhead{HeII (4686 \AA)} 				& \colhead{SiII (6347 \AA)}	\\
		\colhead{}			& \colhead{(E.W.)Flux\tablenotemark{a}\tablenotemark{b}} 			& \colhead{(E.W.)Flux\tablenotemark{a}\tablenotemark{b}} 				& \colhead{(E.W.)Flux\tablenotemark{a}\tablenotemark{b}} 				& \colhead{(E.W.)Flux\tablenotemark{a}\tablenotemark{b}}		& \colhead{(E.W.)Flux\tablenotemark{a}\tablenotemark{b}}}
\startdata
		CSS1111(2012,LBT)	&($-57\pm3$) $1.75\pm0.05$		&($-24\pm2$) $1.35\pm0.05$			&($-28\pm2$) $1.24\pm0.07$			&($-5\pm1$) $0.3\pm0.1$		&($-5\pm1$) $0.17\pm0.03$\\
		CSS1111(2014, APO)	&($-80\pm3$) $1.65\pm0.05$		&($-35\pm3$) $1.6\pm0.1$				&-						&($-11\pm1$) $0.6\pm0.1$		&($<-6$) $<0.1$\\
		CSS1111(2014,LBT)	&($-87\pm3$) $2.47\pm0.02$		&($-34\pm2$) $2.05\pm0.05$			&($-56\pm2$) $1.95\pm0.05$			&($-11\pm1$) $0.65\pm0.05$	&($-10\pm3$) $0.28\pm0.05$\\	
		V418 Ser				&($-44\pm3$) $0.65\pm0.05$		&($-15\pm3$) $0.44\pm0.06$			&($-33\pm3$) $0.53\pm0.04$			&($-4\pm1$) $0.13\pm0.03$	&($<-5$) $<0.08$\\
\enddata
\tablenotetext{a}{Flux measured in $10^{-15}$ erg cm$^{-2}$ s$^{-1}$}
\tablenotetext{b}{E.W. measured in \AA}
\end{deluxetable}

\begin{figure}
\epsscale{1}
\plottwo{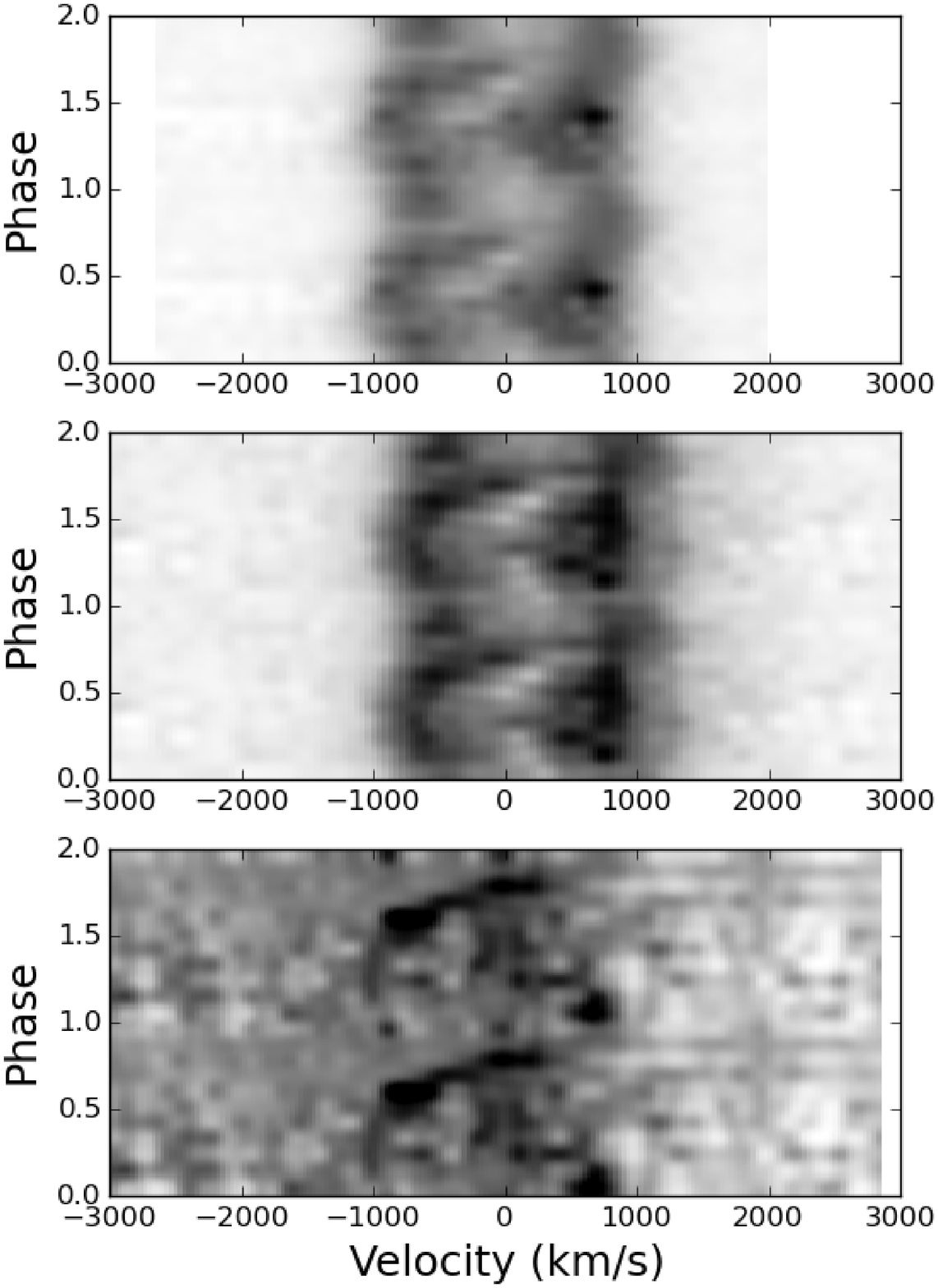}{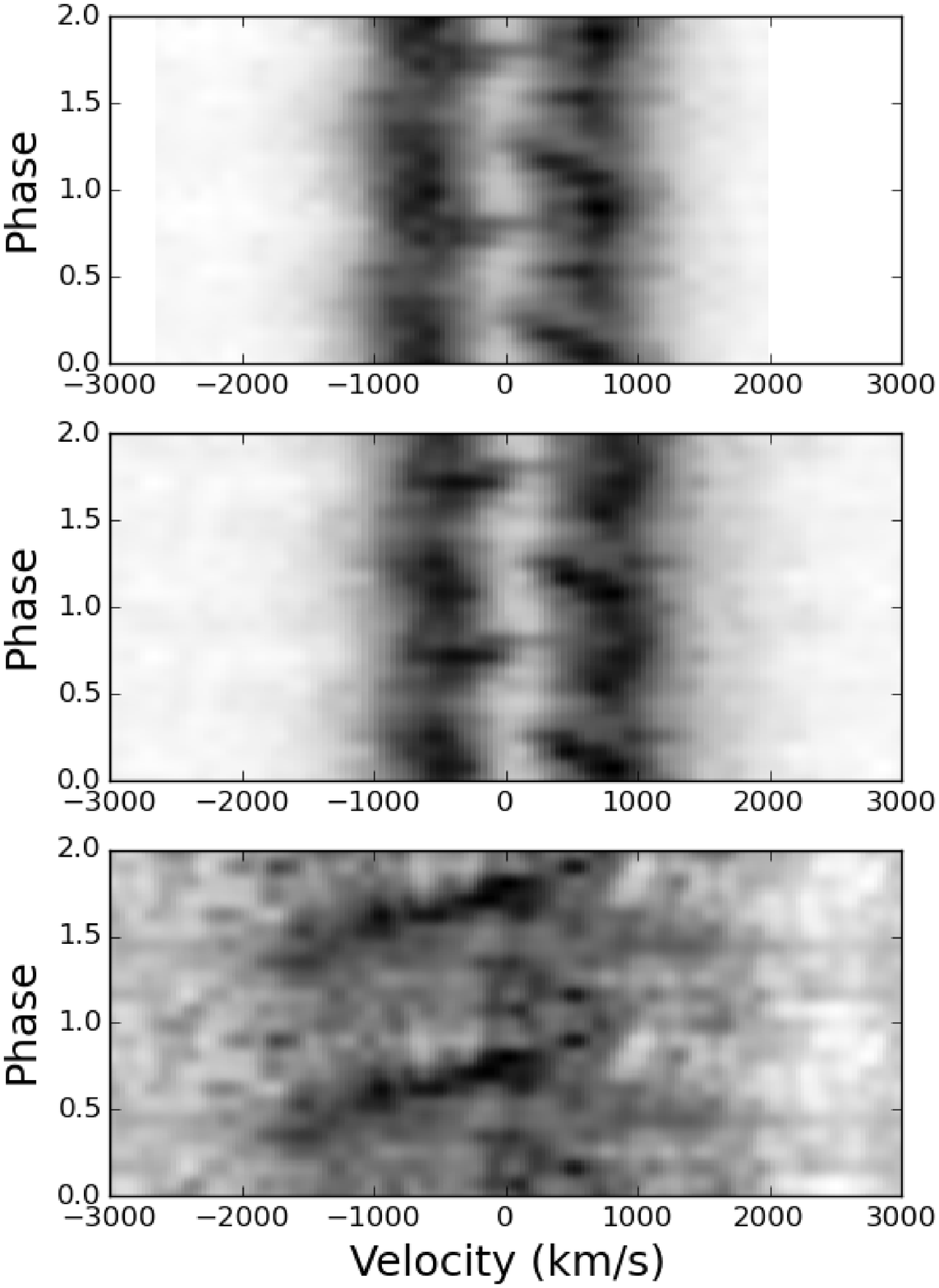}
\caption{The left column shows, in descending order, the H$\alpha$, HeI$_{\lambda5876}$ and the HeII$_{\lambda4686}$ lines in 2012 ,while the right column shows the same for the 2014 data. Both the H$\alpha$ and HeI$_{\lambda5876}$ lines display a complex structure in both years of data, which we interpret as the intertwining of 2 S-wave patterns from different regions of emission within the disk, while the HeII line seems to only appear from phase 0.5 to 0.85 in 2012, and from 0.3 to 0.85 in 2014. The line appears in the blue and trails towards the red as the phase increases, suggesting a strong phase dependence in this line. This dependance is possibly what caused the apparent eclipse seen in \cite{Carter2013}.\label{trailed_spec}}
\end{figure}

Figure \ref{trailed_spec} shows the trailed spectra of the H$\alpha$, HeI$_{\lambda5876}$ and HeII$_{\lambda4686}$ lines in both 2012 and 2014. The continuum around each line was fit with a second order polynomial and then subtracted off each spectrum to increase the signal to noise, and the spectra were Gaussian smoothed to help identify features. The HeII$_{\lambda4686}$ also lies very close to the HeI$_{\lambda4713}$ line. In an attempt to remove contamination of the HeII line by HeI, a model HeI line was constructed using the HeI$_{\lambda5876}$ line. This model was then shifted to  4713 \AA$\:$ and subtracted off. The remaining signal was taken to be the HeII$_{\lambda4686}$ line, and can be seen in the bottom row of Figure \ref{trailed_spec}.

\begin{figure}
\epsscale{1}
\plotone{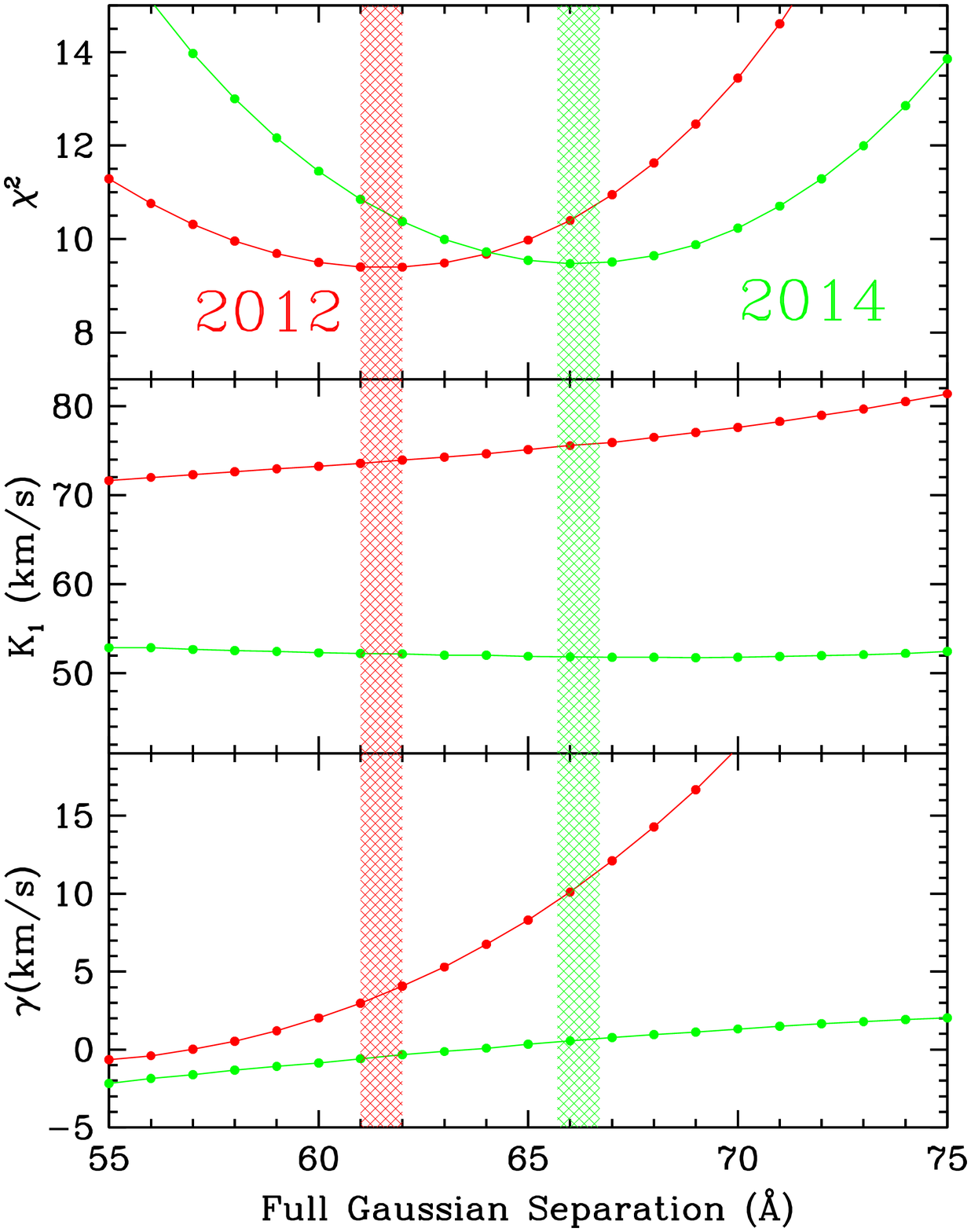}
\caption{Shafter method applied to H$\alpha$ for both the data from 2012 and 2014. The orbital period was fixed to the photometric orbital period of 55.36 minutes. The hatched lines represent the 1$\sigma$ errors for both data sets.\label{Shafter}}
\end{figure}

\begin{figure}
\epsscale{0.8}
\plotone{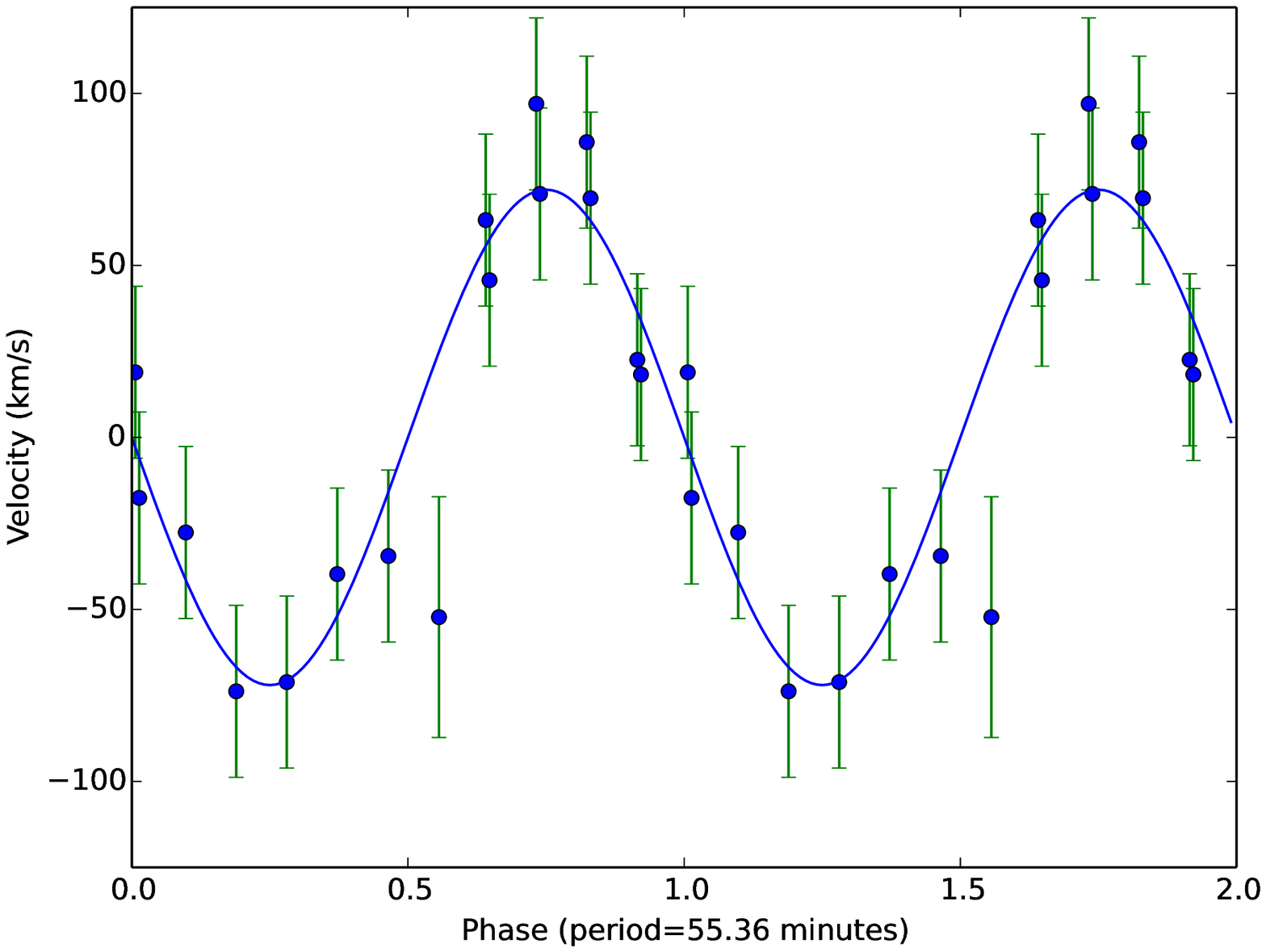}
\plotone{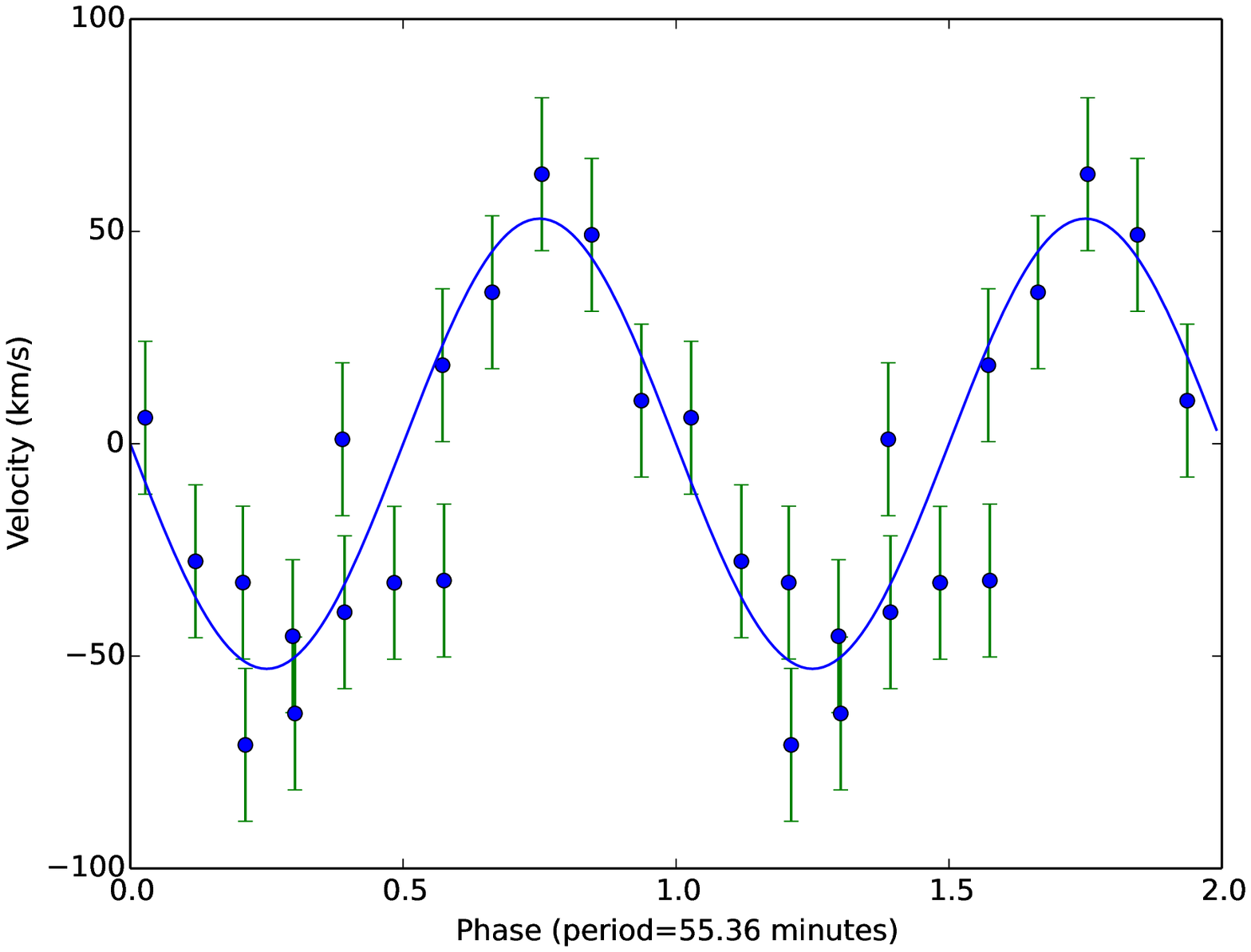}
\caption{The top graph shows the RV curve for the H$\alpha$ line from 2012, which has a $K_{1}=72\:km\:s^{-1}$. The bottom graph shows the RV curve for the 2014 data, which has a $K_{1}=53\:km\:s^{-1}$. \label{RV_curves}}
\end{figure}

\cite{Littlefield2013} used the Shafter method \citep{Shafter1983} on the H$\alpha$ emission line to estimate the orbital period, semi-amplitude velocity of the emission lines and systematic velocity of the system. The Shafter method works by fitting the wings of the emission lines, and since the lines appeared broader in 2014 than in 2012, we expected different results from the fitting. We set the period of the system to be 55.36 minutes, which is the photometric period found by \cite{Littlefield2013} and \cite{Kato2013}, and chose the $K_{1}$ and $\gamma$ which provided the best $\chi^{2}$ to construct a radial velocity using the H$\alpha$ emission line. The results of the Shafter method can be seen in Figure \ref{Shafter}, while the RV curve for the 2014 and 2012 data can be seen in Figure \ref{RV_curves}. We also used the results of the Shafter method to compute the ephemeris of both data sets, such that the zero phase corresponds to inferior conjunction of the companion.

The ephemeris for each data set was found to be:
\begin{equation}
\label{2014eph}
T = 2456771.7776(4) + 0.03845(2)E\:\: (2014)
\end{equation}
\begin{equation}
T = 2456094.6784(4) + 0.03845(2)E\:\: (2012)
\end{equation}

\begin{figure}
\epsscale{0.8}
\centering
\plottwo{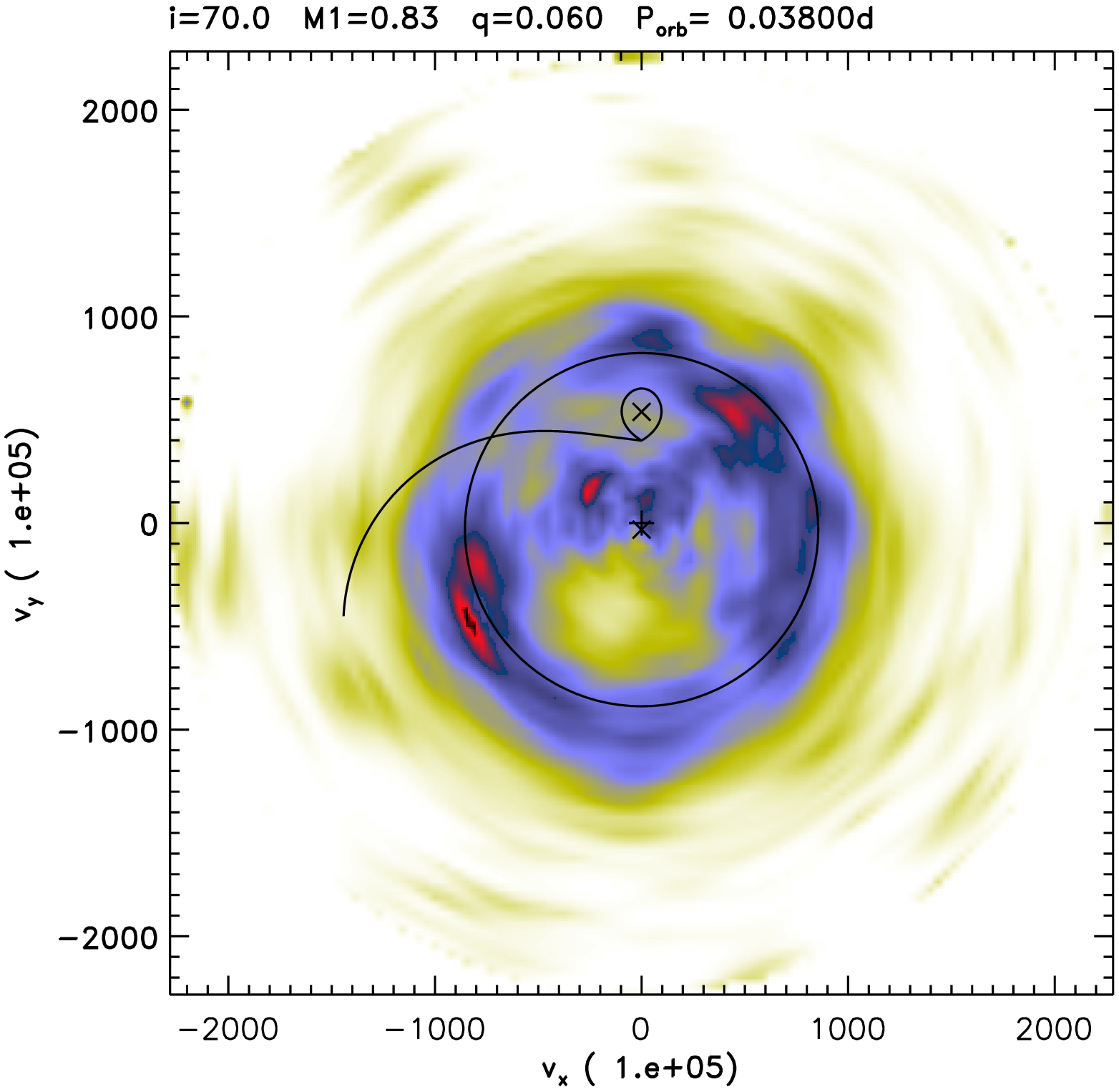}{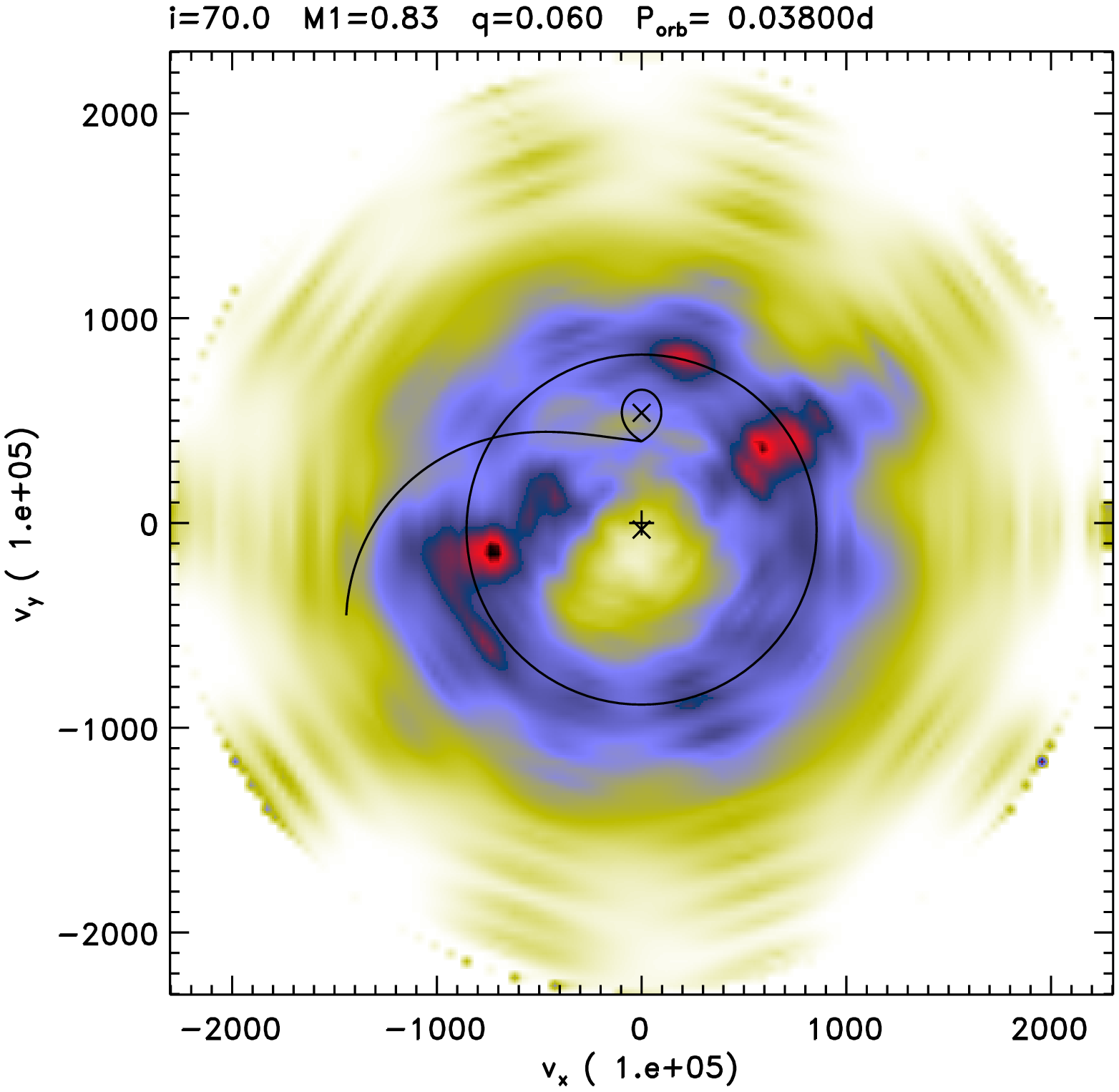} 
\newline
\newline
\plottwo{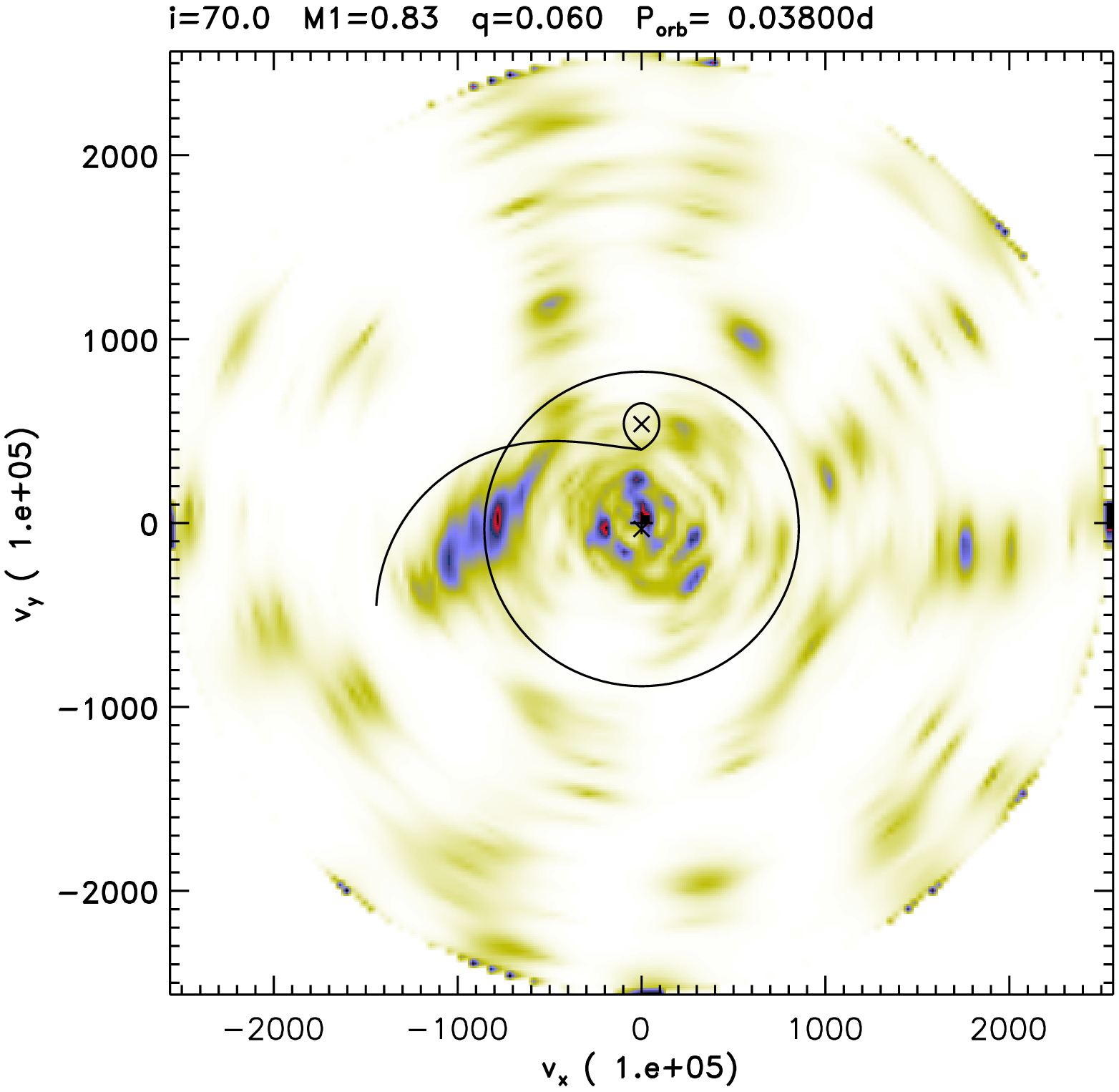}{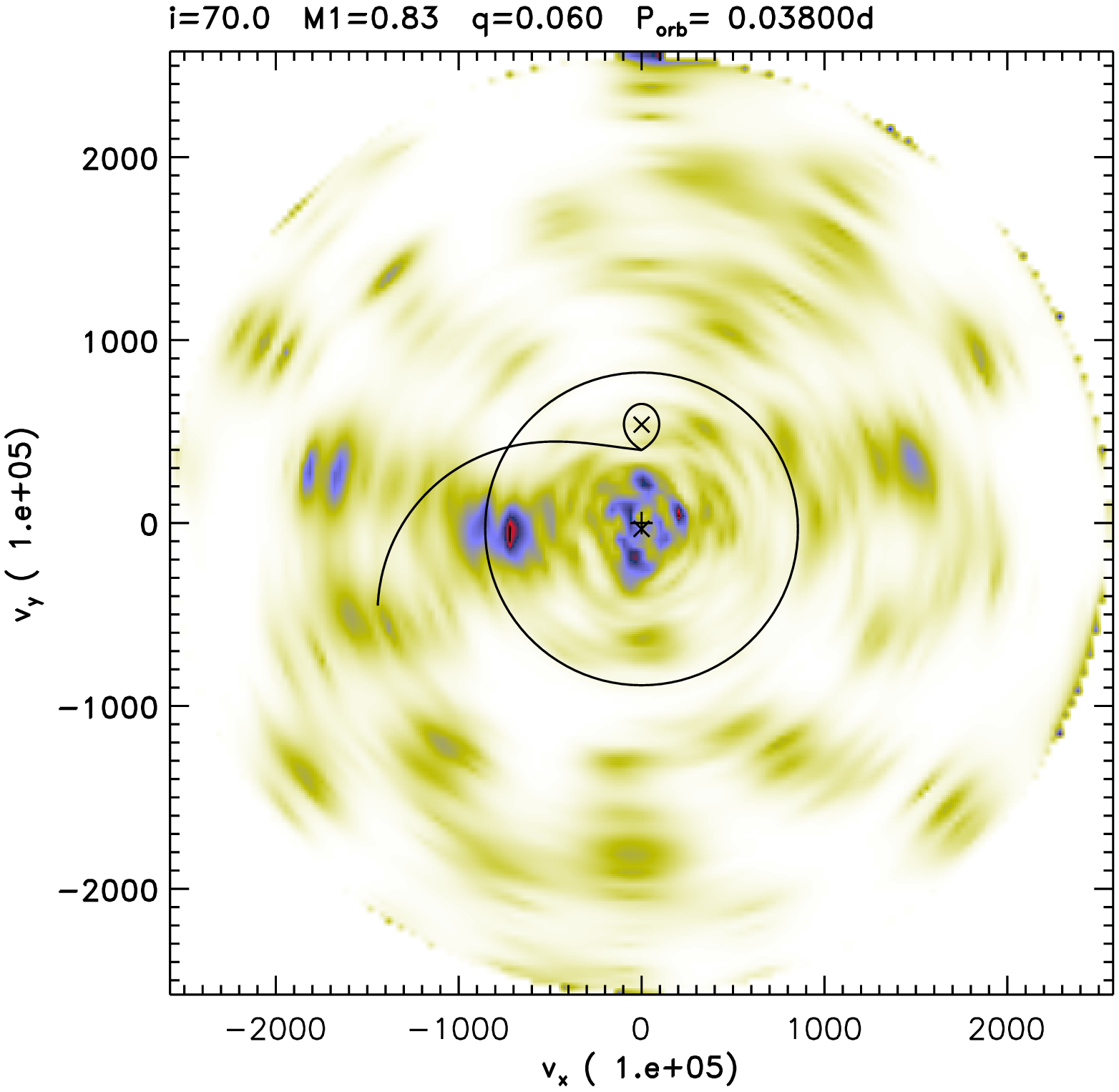}
\caption{The left column shows the doppler maps generated from the 2012 LBT spectra, while the right column shows the map from the 2014 spectra. The top row are the H$\alpha$ maps, and the bottom are the HeII$_{\lambda4686}$ maps. In the H$\alpha$ maps, the main features remain distinct between the maps: the area of enhanced emission in the lower left quadrant and the upper right quadrant, which are associated with a tidally induced spiral shock \citep{Littlefield2013}. The solid circle represents the 2:1 orbital resonance velocity of 750 km s$^{-1}$. In the He II maps, emission is only seen coming from a single point, located at a phase of 0.75, and is slightly offset from the main region of emission in the lower left quadrant seen in the H$\alpha$ maps. We adopted a white dwarf mass of 0.83 M$_{\odot}$, in line with \cite{Littlefield2013}.}\label{DopplerMaps}
\end{figure}

We also constructed Doppler tomograms \citep{Marsh1988} using the code in \cite{Spruit1998}. The tomograms of H$\alpha$ were used to see if the structures present in 2012 were still evident in 2014 and those of HeII$_{\lambda4686}$ were used to check for the phase dependence of the HeII emission. The Doppler maps can be seen in Figure \ref{DopplerMaps}. The HeII emission is coming from a single point seen in both 2012 and 2014, and is slightly offset from the area of H$\alpha$ emission in the lower left quadrant. It is obvious from direct comparison of the H$\alpha$ Doppler maps that the disk in 2014 extended out to higher velocities. The two areas of enhanced emission proposed by \cite{Littlefield2013}, thought to be the accretion stream and a spiral shock located at the 2:1 resonance \citep{Lin1979} were still present in the 2014 map. However, the phasing between our 2012 tomogram and the tomogram presented by \cite{Littlefield2013} is different. \cite{Littlefield2013} used the maximum of the optical light curve, which also corresponds to the maximum of the radial velocity curve, to define phase 0, while here, the 0 point of the radial velocity is used, corresponding to a phase shift of 0.25 between the tomogram presented in \cite{Littlefield2013} and the tomograms presented here. This difference in the definition of the 0 phase leads to different interpretations of the Doppler tomograms. 

\cite{Littlefield2013} interpreted one of the areas of emission in the 2012 tomograms as being the traditional hot spot, as it appeared at the expected phase in the tomogram) the (-V$_{x}$,+V$_{y}$) quadrant), and the other emission region as being a single spiral arm. However, using our 0 phase, which is well defined by the radial velocity curves as inferior conjunction of the companion, we find that neither of the emission regions lies close to the expected phase of the hot spot in 2012 or 2014. Instead, it is likely that both emission regions are spiral arms, and that there is no strong emission coming from the intersection of the accretion stream with the disk. \cite{Littlefield2013} also suggested that the emission feature in their tomogram not associated with their hot spot could be due to accretion stream overflow, where material striking the disk at the hot spot overflows and strikes the disk closer to the WD(\citealt{Armitage1996}; \citealt{Armitage1998}). The velocity of this second hot spot should be higher than the original hotspot, as the overflow material is accelerated to higher velocities by the WDs strong gravitational field \cite{Armitage1998}. Since both emission regions in our tomograms are orbiting with very similar velocities, we find this scenario very unlikely, supporting the tidally induced spiral shock proposed by \cite{Littlefield2013}.

\subsection{UV Spectrum}
\subsubsection{Average Spectrum}
\begin{figure}
\epsscale{0.8}
\plotone{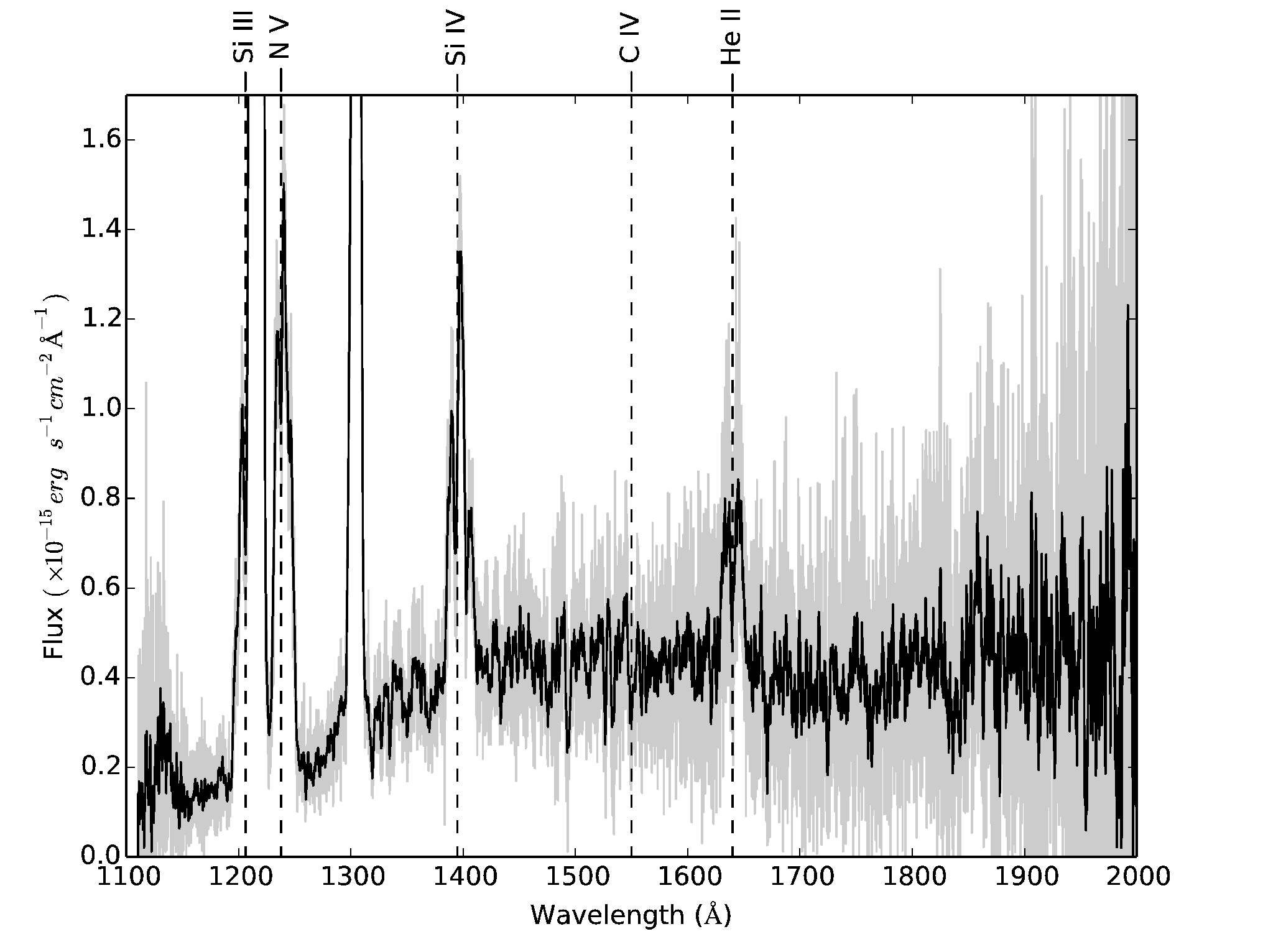}
\plotone{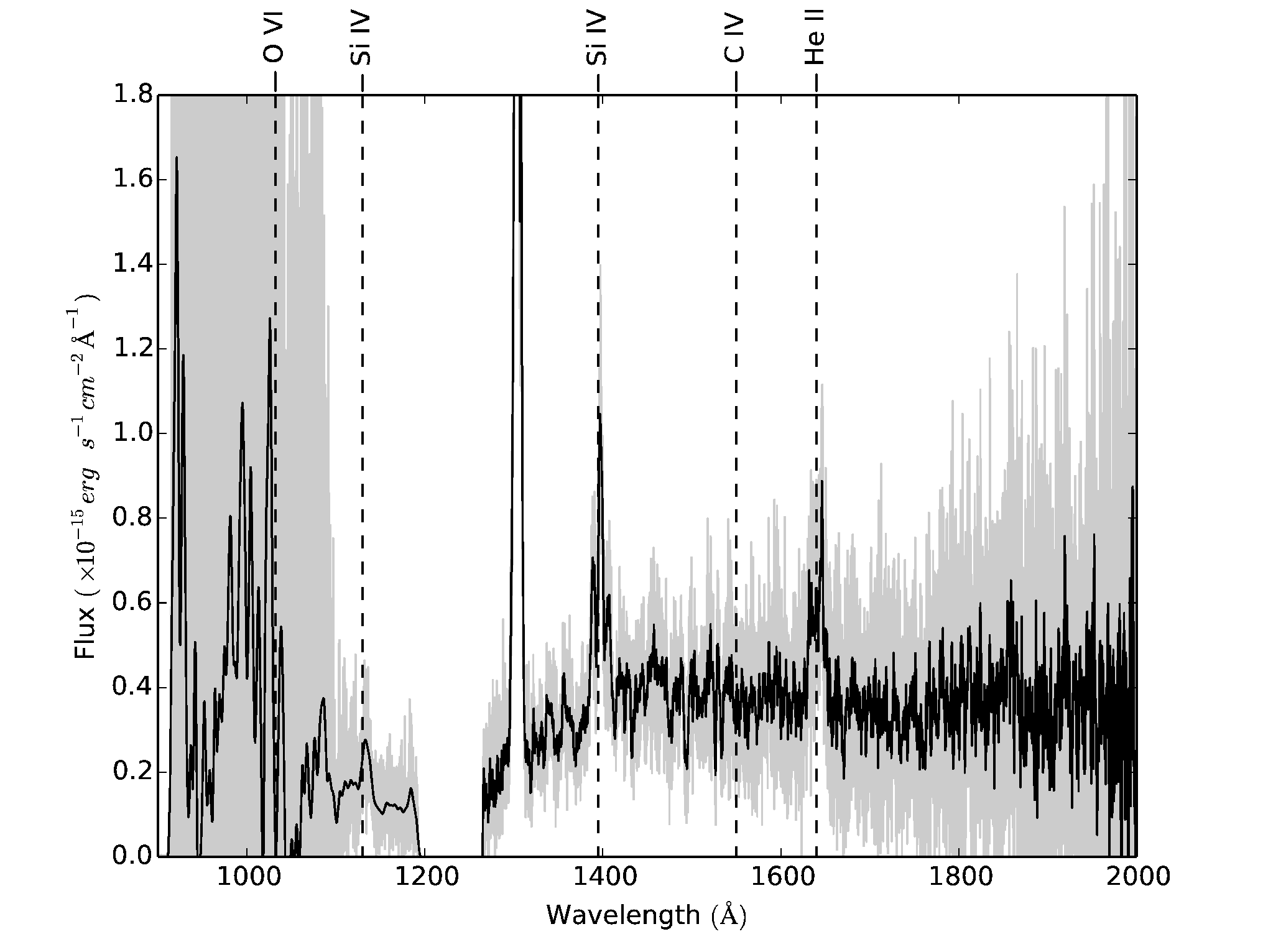}
\caption{The average UV spectrum from the COS instrument aboard the HST. The top panel shows the average spectrum using the G140L grating in the 1105\AA$\:$ setting. The spectrum has been boxcar smoothed with a boxcar smoothing width of 1.2\AA$\:$. The bottom panel shows the spectrum using the G140L grating in the offset position. The blue part of the spectrum was smoothed by convolution with a gaussian with $\sigma=0.8$\AA$\:$ in an attempt to clean up the noise. The red part of the spectrum was boxcar smoothed in the same way as with the top figure. The grey in each figure shows the unsmoothed averaged spectrum.}\label{UV_Spec}
\end{figure}

\begin{figure}
\epsscale{1}
\plotone{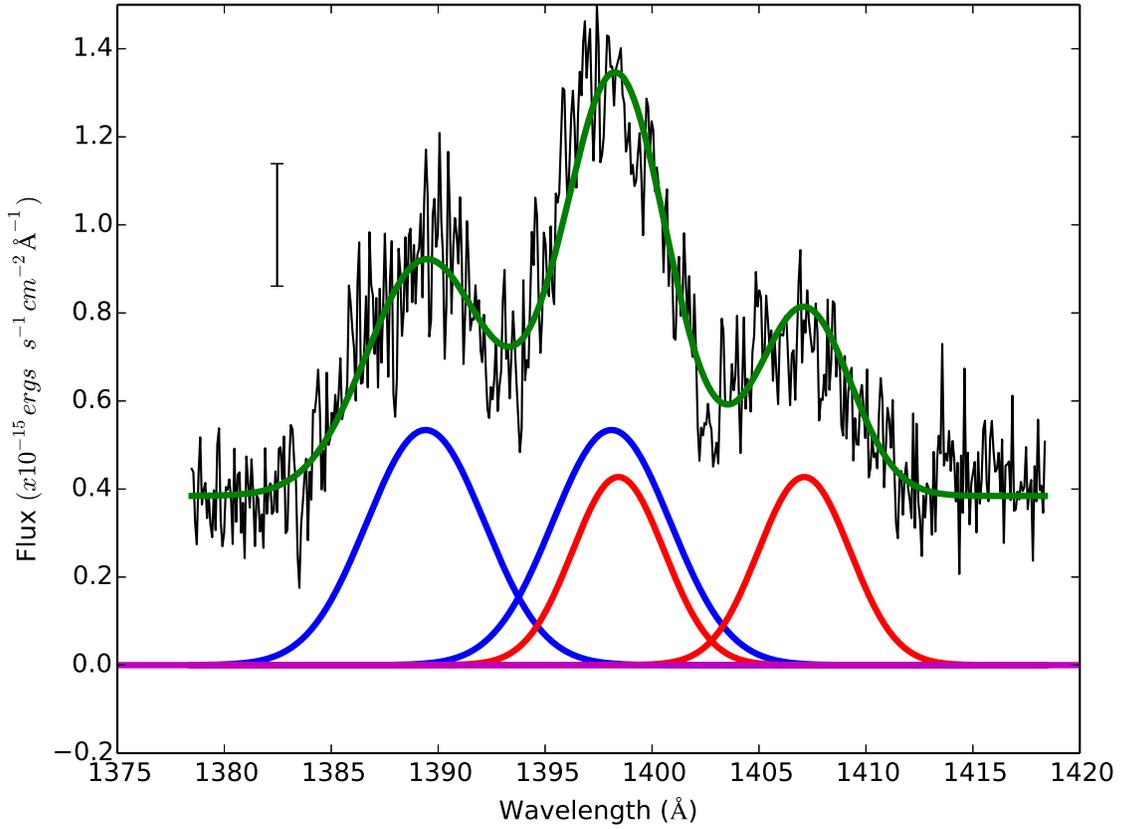}
\caption{An example of how we explain the triple peaked features in the UV spectrum. Si IV has a doublet at 1393\AA$\:$ and 1402\AA$\:$. Each doublet is doppler shifted due to the source of the emission being the accretion disk. We modelled each doublet as a pair of gaussians, and the resulting total structure looks triple peaked. The single bar shows the typical size of an error bar in the spectrum}\label{SiIVline}
\end{figure}

The average UV spectrum for both grating positions taken using the COS instrument can be seen in Figure \ref{UV_Spec}. The most readily identifiable lines in the spectrum are NV$_{\lambda1240}$, SiIV$_{\lambda1400}$ and HeII$_{\lambda1640}$. Both the SiIV and NV lines are doublets, and the blending of these doublets is attributed to the triple peaked appearance of these lines. This is more readily seen in Figure \ref{SiIVline}, which shows how the SiIV doublet is split and makes a triple peaked feature. The peak-to-peak separation of the lines in the doublet was found to be $8.8\pm0.6$ \AA$\:$, which gives a velocity of $950\pm60$ km s$^{-1}$, which suggests the SiIV emission is coming from an inner, high excitation region of the disk.  The details of these 3 lines can be seen in Table \ref{uv_ew}, along with an upper limit on the CIV line. It is important to note the distinct lack of carbon emission in this spectrum, which will be addressed in Section \ref{EMS_Sect}. There also is no clear evidence for the O VI$_{\lambda1032}$ line. However, it is probable that this is due to poor signal-to-noise in this region of the spectrum, as shown by the unbinned spectrum. The OI$_{\lambda1300}$ line is air glow, and is not attributed to CSS1111.

\begin{table}
	\begin{center}
	\caption{Fluxes of some of the most prominent emission lines in the UV spectrum\label{uv_ew}}
	\begin{tabular}{cc}
		\tableline\tableline
		Line & Flux\\
		& $10^{-15}$ erg cm$^{-2}$ s$^{-1}$\\
		\tableline
		NV&16$\pm$3\\
		SiIV&12$\pm$2\\
		CIV&$<$5\\
		HeII&6$\pm$3\\
		\tableline
	\end{tabular}
	\end{center}
\end{table}

\subsubsection{TIME-TAG Light Curves}
\begin{figure}
\epsscale{0.8}
\plotone{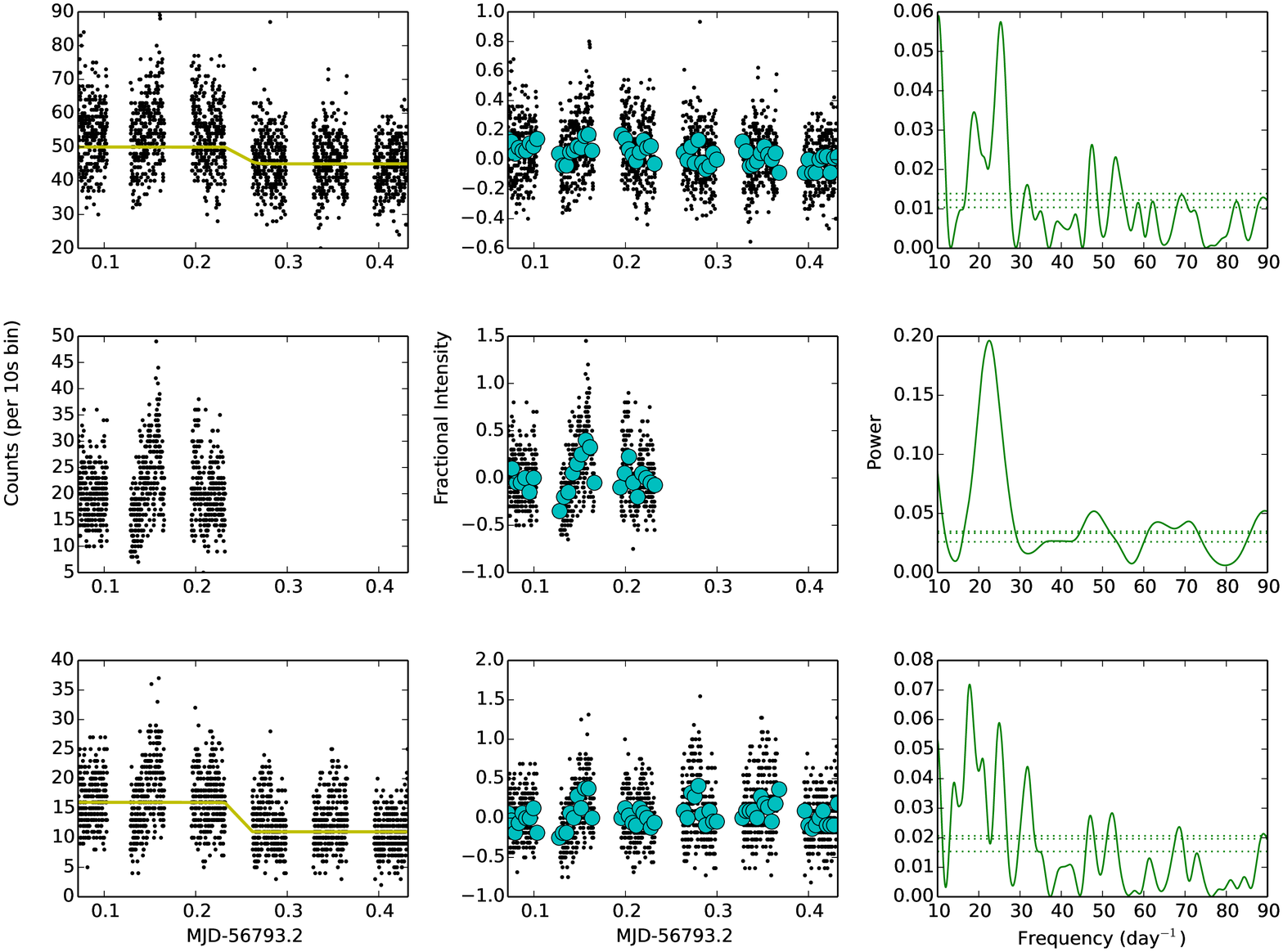}
\epsscale{0.6}
\plotone{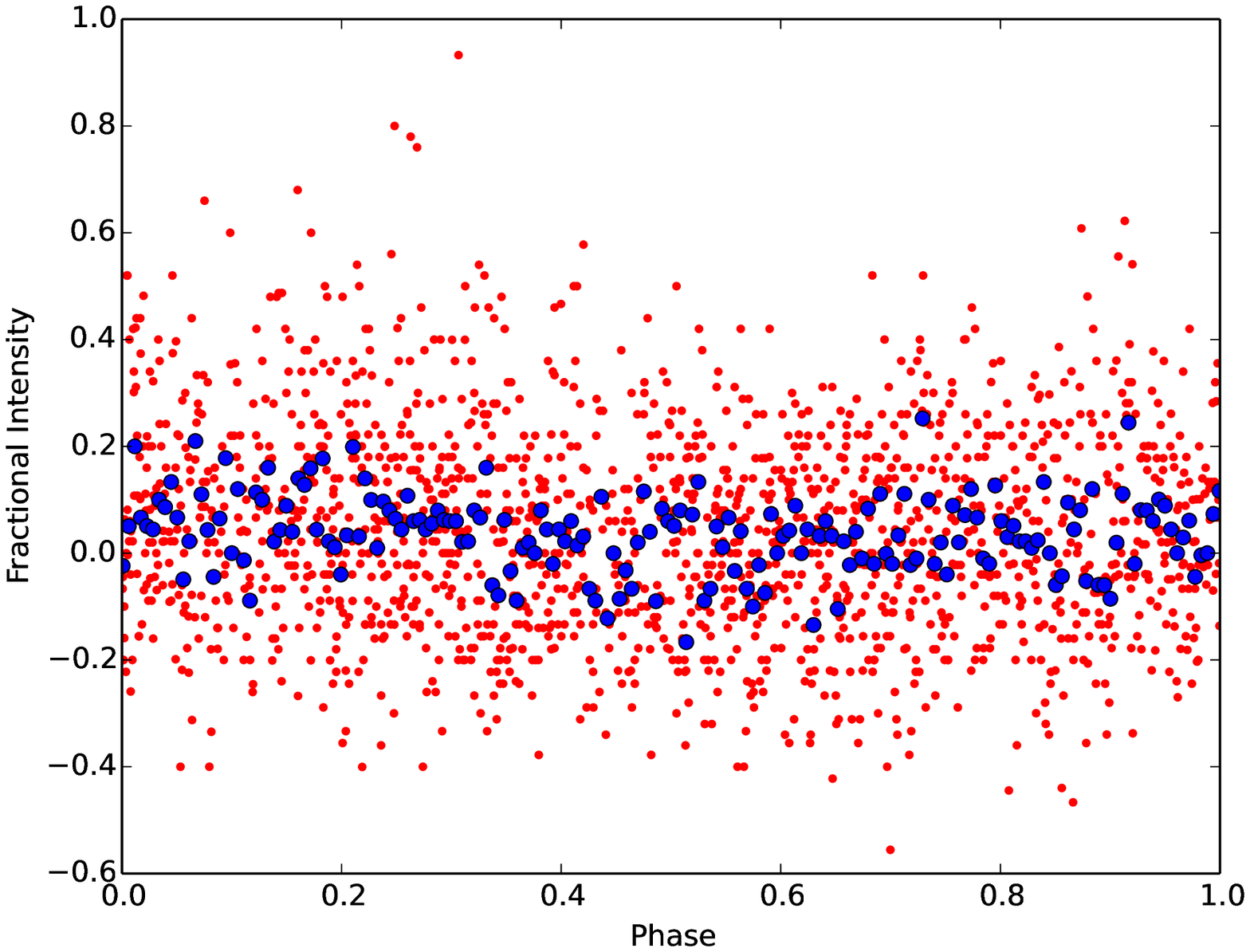}
\caption{ The time tag data extracted from the HST COS data. 3 different wavelength segments were analysed: 1327-1379 \AA$\:$ and 1412-1610 \AA$\:$ as a sample of the continuum (top row); 1226-1255 \AA$\:$ to look at the N V feature (middle row); 1379-1412 \AA$\:$ to look at the Si IV feature (bottom row). The left column shows the background subtracted counts for each segment, while the yellow line in the continuum and Si IV plots show the long term trend due to shifting the G140L grating, which was then subtracted. The middle column shows the fractional variation after the background had been subtracted from the counts and the light curve had been normalised by the long term trend. The right column shows the results of the Lomb Scargle Periodogram. The lines represent the 95\%, 99\% and 99.5\% confidence levels, found by using a bootstrap Lomb Scargle Periodogram routinue with 1000 iterations. The bottom graph shows the continuum light curve phased tp the 57 minutes period, using the ephemeris in equation \ref{2014eph}. The red points are the corrected, normalised counts, while the blue points are a binned light curve with a bin width of 10 data points.\label{COSTTLC}}
\end{figure}

The TIME-TAG spectroscopic data from HST was used to make light curves for three different wavelength ranges to look for variability of the continuum or emission lines. The three ranges selected were: 1327-1379 \AA$\:$ and 1412-1610 \AA$\:$ for the continuum, $1226$\AA$\: - 1255$\AA$\:$ for the NV emission feature and $1379$\AA$\: - 1412$\AA$\:$ for the SiIV emission feature. The light curves show no obvious periodic features other than the orbit, and no eclipse is visible. The background subtracted light curves, binned up to 10s temporal resolution, can be seen in the middle column of Figure \ref{COSTTLC}. However, \cite{Carter2013} noted that the grazing eclipse seen in the HeII$_{\lambda4686}$ line may only occur during outburst, when the disk radius is greater than during quiescence. Since our observations were taken when the system was in quiescence, the non-detection of an eclipse does not rule out the eclipse seen by \cite{Carter2013}. 

The data were subjected to a Lomb Scargle periodogram (\citealt{Lomb76}; \citealt{scargle82}) using the astroML python library \citep{astroMl}. The power spectrum for each of the 3 light curves can be seen in the right column of Figure \ref{COSTTLC}, where the confidence levels were obtained by using the Lomb Scargle bootstrap routine over 1000 iterations. The continuum returned a strong peak at $25\pm1$ cycles/day, ($57\pm3$ min), while the Si IV range returned a strong peak at $25\pm2$ cycles/day ($57\pm5$ min). The N V wavelength range, which was only visible for 3 orbits due to the shift in the grating, has a peak at $22\pm4$ ($65^{+15}_{-10}$ min).

\section{White Dwarf Temperature}
\begin{figure}
\epsscale{1}
\plotone{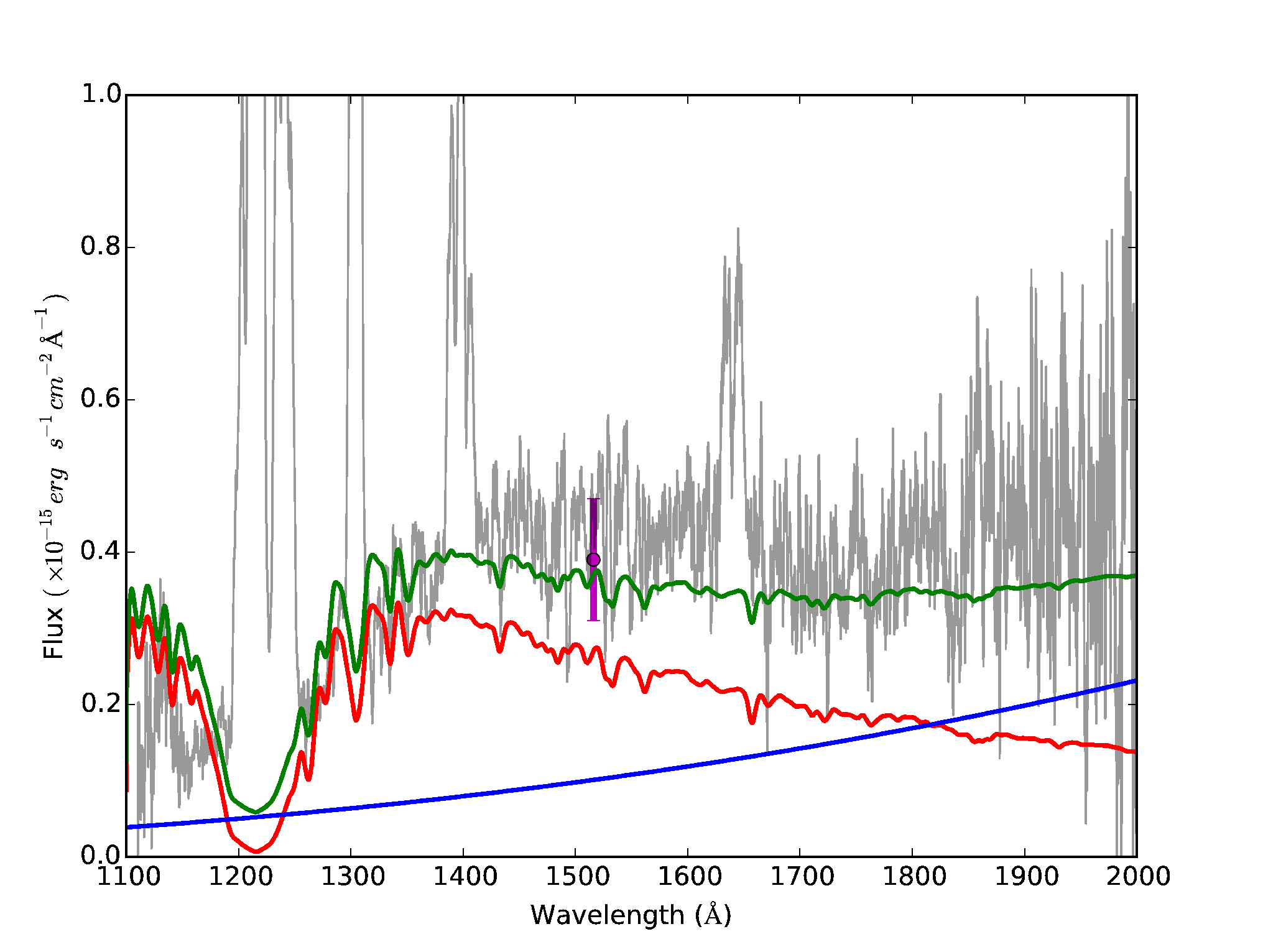}
\caption{The best fit model to the UV spectrum (shown in grey) using TLUSTY and a power law contribution from an accretion disk. The red line represents the WD model for $T=16000 K$ and log $g=8$, while the blue line shows the power law contribution from the disk with $\alpha=5$. The green line is the combination of both these models. The purple point is the GALEX FUV (effective wavelength = 1516 \AA) flux of 0.39$\pm$0.08 erg s$^{-1}$ cm$^{-2}$ \AA$^{-1}$.\label{UV_fit}}
\end{figure}

In order to obtain an estimate for the temperature of the WD, we used Hubeny's TLUSTY, SYNSPEC and ROTIN programs (\citealt{Hubeny1988}; \citealt{Hubeny1995}) to generate synthetic spectra of white dwarfs with various temperatures and log $g$. To account for contribution from the accretion disk in the UV, we adopt a power law of the form
\begin{equation}
f_{\lambda} = f_{\lambda}(1200\text{\AA})\times\left(\frac{\lambda}{1200 \text{\AA}}\right)^{\alpha-2}
\end{equation}
as done in \cite{Long2009}. Here, $f_{\lambda}(1200$\AA$)$ is the disk contribution to the overall flux at 1200 \AA$\:$ which was taken to be 10\% of the continuum around this point. It is important to note that when fitting a white dwarf and power law model to a spectrum, there is a strong covariance between the temperature of the WD model and the $\alpha$ of the power law. In this case, the shape of the spectrum at low wavelengths ($<$1400 \AA) reveals important information about the WD spectrum due to the weak contribution of the disk, and the strength of the Lyman $\alpha$ absorption by the WD can also help give important constraints on the temperature of the WD. Taking into account the shape of the spectrum, the width of the Lyman $\alpha$ absorption and the unusual structure of this disk, our best fit model can be seen in Figure \ref{UV_fit}. The WD temperature was found to be $T=16000\pm2000 K$, with $\alpha=5\pm1$.

\cite{Carter2013} found the WD temperature in CSS1111 to be 12000$\pm$1000 K by modelling the optical spectrum along with the GALEX FUV and NUV fluxes of 30.74$\pm$5.92 $\mu$Jy and 48.25$\pm$1.34 $\mu$Jy respectively. However, they did not include any UV contribution from the accretion disk when modelling the system. Since the GALEX fluxes are broadband, it is difficult to get an accurate SED to to fit for the temperature. Our model is consistent with the GALEX FUV measurement, as shown in Figure \ref{UV_fit}. Extrapolating our UV WD temperature to the optical yields a flux from the WD below the observed optical flux, which is expected as the accretion disk should be dominant in the optical.

\section{An evolving disk}
\label{EvolvingDisk}
The two Doppler maps in Figure \ref{DopplerMaps} and the averaged spectra in Figure \ref{average_spec} show a significant change has occurred in the 2 year gap between LBT observations. In 2014, there was significant emission in the disk for velocities above 1300 km s$^{-1}$, while in 2012 there was very little. Also, the equivalent width of the optical emission lines increased.

\begin{figure}
\epsscale{1}
\plotone{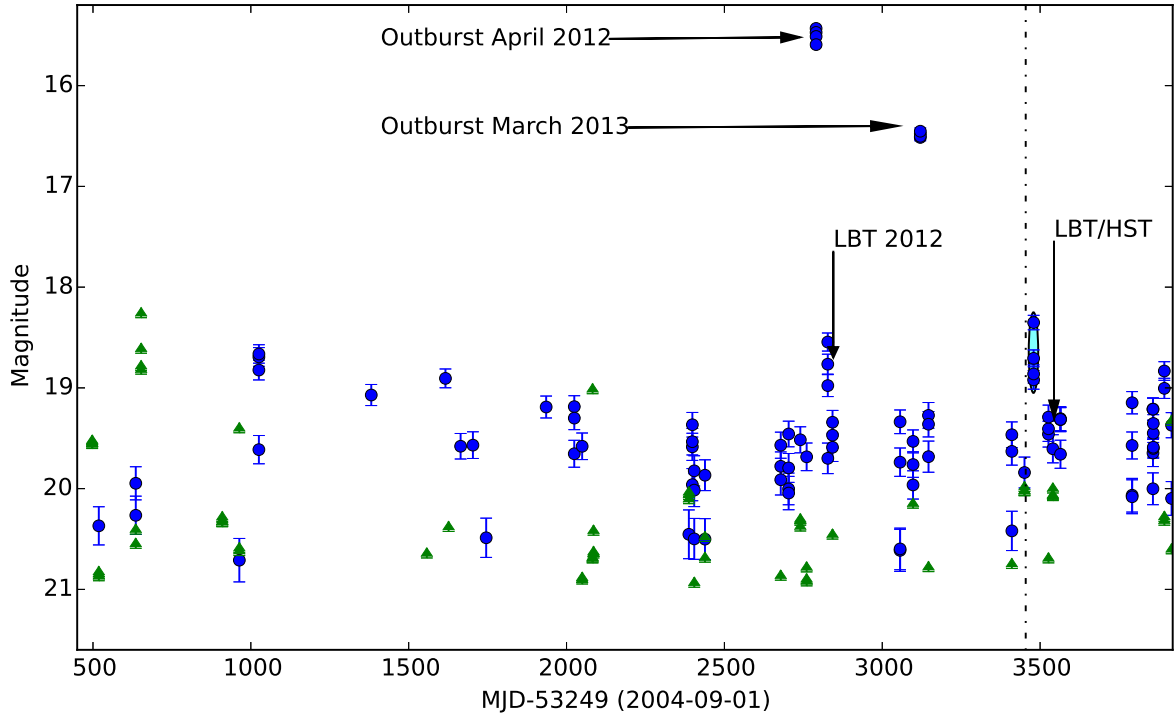}
\caption{The longterm light curve from the Catalina Surveys Data Release 2 (blue for detections, green arrows for lower limits). The data starts in 2006, shows 2 outbursts, one in April 2012 and one in March 2013, and then ends in April 2015. It looks like there was no outburst between March 2013 and our LBT and HST observations in April 2014, however the gap in observations from April 2013 to January 2014 makes this difficult to confirm. There is a possible outburst seen by the CRTS during March 2014. The data taken on 2014-03-11 have a magnitude of 18.5-19, which is thought to be the end of an outburst and is hightlighted, while the quiescent magnitude of CSS1111 is 19.5. The dashed black line here represents when an outburst, which lasts for 25 days \citep{Kato2013}, would have had to occur for the system to be at 18.5 on 2014-03-11. Since the CRTS only samples a region of the sky every 28 days, it could have missed the system at its brightest and caught only the tail end of the outburst. The light curve also shows periods of non detection, where the source seems to be fainter than 20.5 mag, which is outside the 0.2 mag orbital variation. This leads us to believe CSS1111 might exhibit a very faint quiescent state.}\label{CRTS}
\end{figure}

It is possible that this is due to an increase in mass in the inner disk at the time of the 2014 observations compared to the 2012 observations. The observations in 2012 were taken only 2 months after the system had gone into outburst. In fact, the system was still above its quiescent magnitude 2 weeks prior to the LBT observations, as the outbursts have a duration of 25 days, as seen in Figure 46 of \cite{Kato2013}. Following the theory for disk instabilities causing the dwarf nova outburst, it would be expected that the disk would be have a lower mass density in observations taken around this time \citep{Osaki1996}. Figure \ref{CRTS} shows the long term light curve of CSS1111 from the Catalina Real-Time Transient Survey (CRTS) Data Release 2 \citep{Drake2009}. The light curve shows the outburst in 2012, along with another outburst in March 2013.

Since March 2013, no outburst has been conclusively detected. This includes a gap in observations between April 2013 and January 2014. However, as seen in Figure \ref{CRTS}, the system seemed to be brighter than its quiescent magnitude on 2014 March 11, 43 days before the LBT observations and 66 days before the HST observations. Since there were no observations made of the system by the CRTS for 28 days up to these observations, and the outburst in 2012 lasted 25 days, then it is likely that the system was in outburst during this time, and the CRTS caught the system as it was returning to quiescence. This means that in 2012, the system only had 14 days to recover before the LBT observations, compared to the 43 days or more it had in 2014. This extended period of time, and the possible high rate of mass accretion in this system, should be enough to explain the differences seen in the optical spectra, as the accretion disk had longer to recover in 2014.

\begin{figure}
\epsscale{1}
\plotone{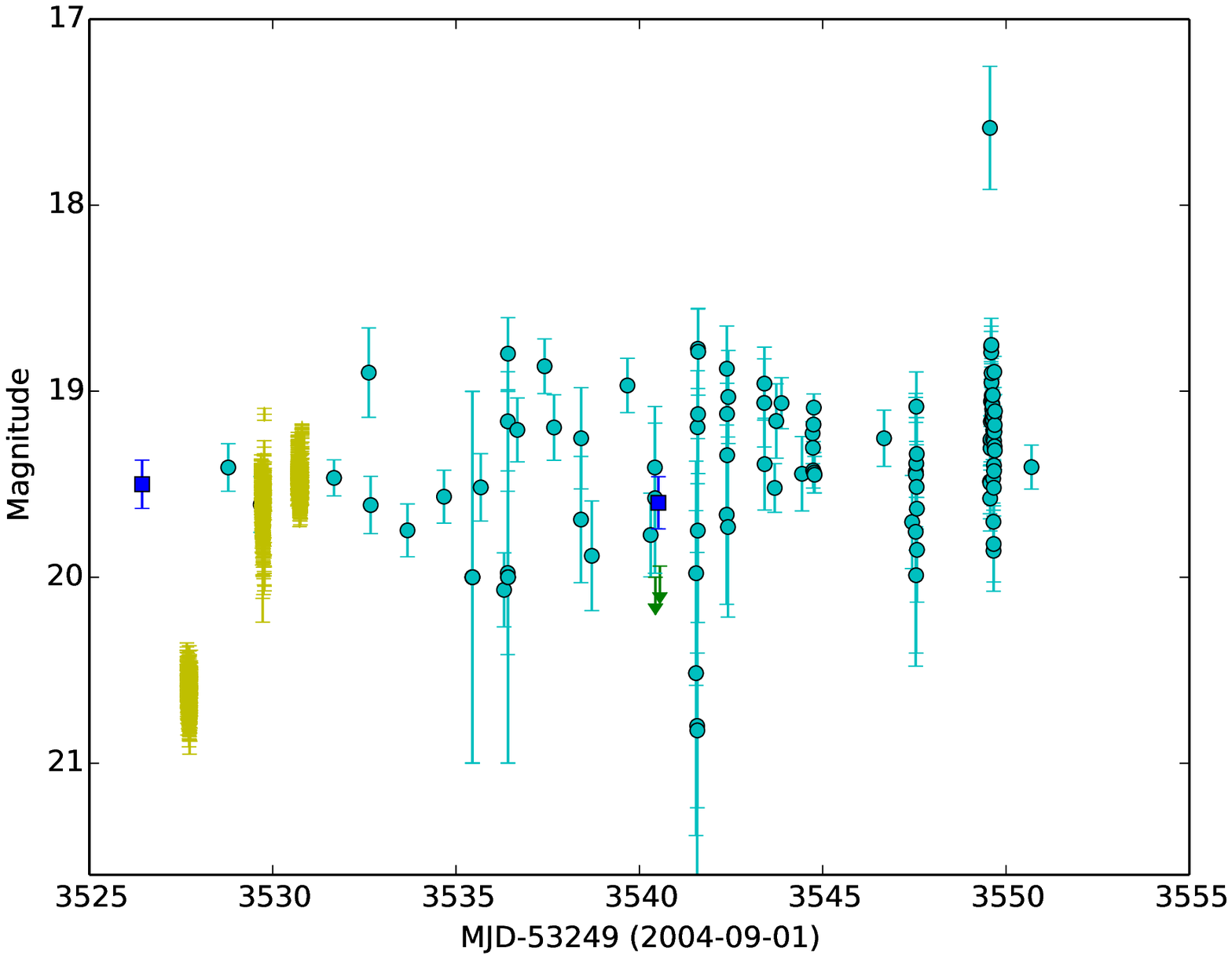}
\caption{The AAVSO observations (cyan), combined with the observations taken from the Catalina Surveys Data Release 2 (same as Figure \ref{CRTS}), around the HST observations. CSS1111 has a variation of 0.2 magnitudes during its orbital period, which accounts for the fluctuations seen in the AAVSO light curve, but not for the very low non detections ($>$ 20.5 magnitudes) seen in the CRTS light curve. The yellow shows the data obtained with the VATT as detailed in Section \ref{VATT}.}\label{AAVSO}
\end{figure}

The AAVSO (Figure \ref{AAVSO}) and CRTS (Figure \ref{CRTS}) light curves show the photometric variability of CSS1111. While the AAVSO light curve finds CSS1111 with a mean magnitude of 19.5, with error bars larger than the expected orbital modulation of 0.2 magnitudes \citep{Littlefield2013}, the CRTS light curve shows very different behaviour - CSS1111 is not detected for periods of time with a limiting magnitude of 20.5 at stages, and often even down to 21 magnitude. The light curve of CSS1111 has been seen to vary by 0.6 magnitudes over a single night (see Figure 1 of \citealt{Littlefield2013}), and combining this single night of observations with the CRTS light curve leads us to believe CSS1111 might exhibit a hitherto unknown deeper quiescent state, where the magnitude of the object drops to below 20.5. This is further confirmed by the VATT observations shown in Figure \ref{AAVSO}, where CSS1111 had magnitude $>20.5$ during the first night of VATT data.

\cite{Littlefield2013} proposed that the region of enhanced emission in the Doppler tomograms not associated with the hot spot is possibly due to a spiral shock in the accretion disk. Typically, spiral shocks are not seen in CVs, as the radius at which the 2:1 orbital resonance ($R_{2:1}$) occurs is beyond the maximum radius of the accretion disk. However, for systems with a small mass ratio, $q<0.1$, $R_{2:1}$ lies within the accretion disk, allowing spiral shocks to form. The estimated mass ratio of CSS1111 is 0.06 \citep{Kato2013}, and as such, $R_{2:1}$ in velocity space lies at 750 km s$^{-1}$. This radius is plotted on both Doppler tomograms, and shows that the emission region on the right hand side of the tomograms lies along this radius. Even with a disruption to the disk, the spiral shock should be stable up to the 3:1 orbital resonance, which lies at a velocity 850 km s$^{-1}$, and the emission region lies within this radius. 

The trailed spectra in Figure \ref{trailed_spec} confirm similar structure in the disk between the observations in 2012 and in 2014. The double S-wave shapes of two emission areas intertwined seen in the 2012 H$\alpha$ and HeI lines is still visible in the 2014 data. Also, the weak HeII$_{\lambda4686}$ line, which is partially blended with the 4713 line, is only visible for part of the orbit in both 2012 and 2014. The shape of the HeII line is an arc that starts with a velocity of $<-1000$ km s$^{-1}$ at phase 0.5 and moves towards 0 km s$^{-1}$ at phase 1.0. 

This shape follows the arcs which move from $<-750$ km s$^{-1}$ to 0 km s$^{-1}$ in the H$\alpha$ and HeI lines very well. We attribute the origin of the HeII emission to the same area which produces these arcs, which is the emission area seen in the lower left quadrant of the Doppler tomograms. This is re-enforced by the Doppler tomograms depicted in Figure \ref{DopplerMaps}, which show the HeII emission coming from a similar area to one of the regions of enhanced emission in H$\alpha$.

\section{The Evolved Main Sequence Channel}
\label{EMS_Sect}
\begin{figure}
\epsscale{1}
\plottwo{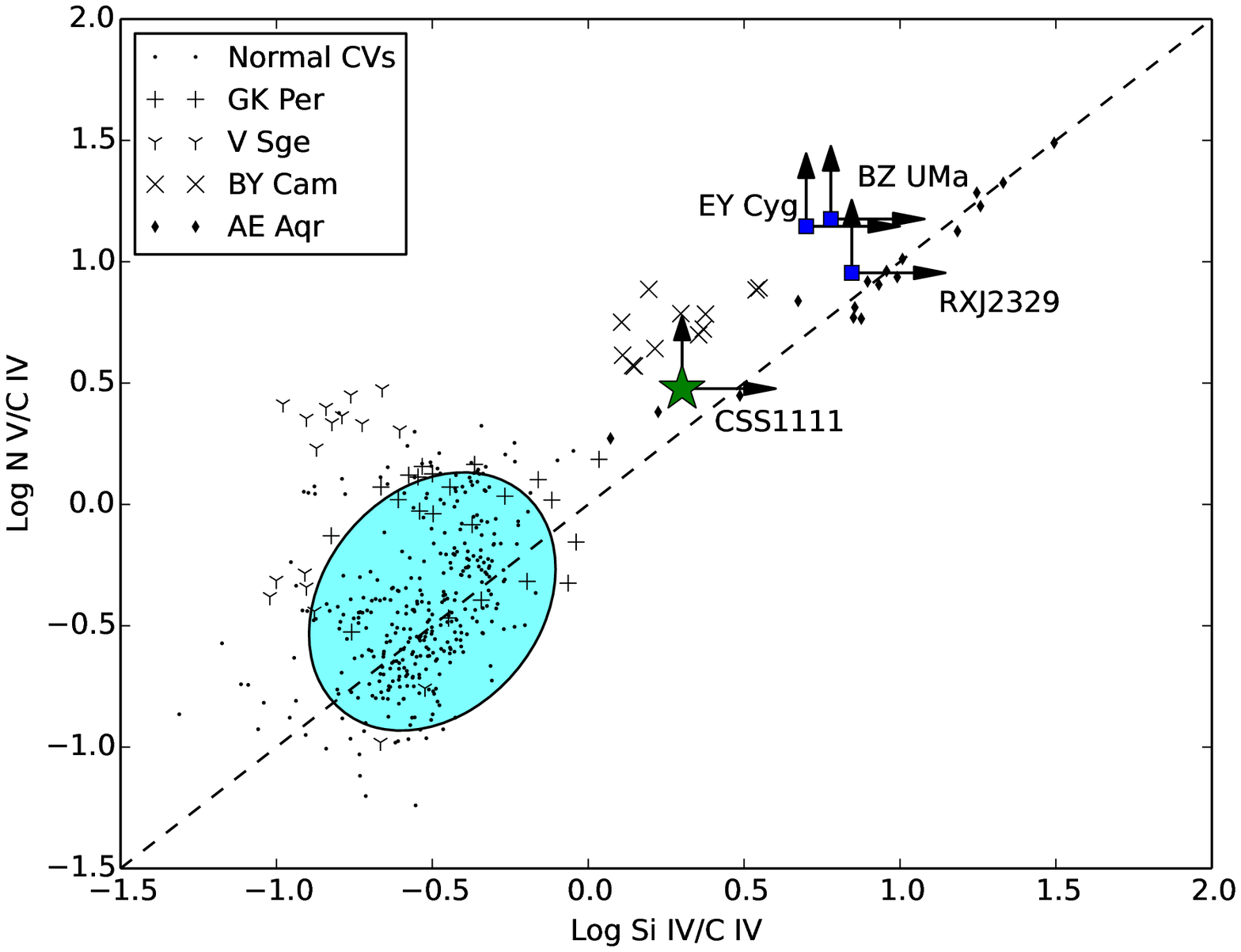}{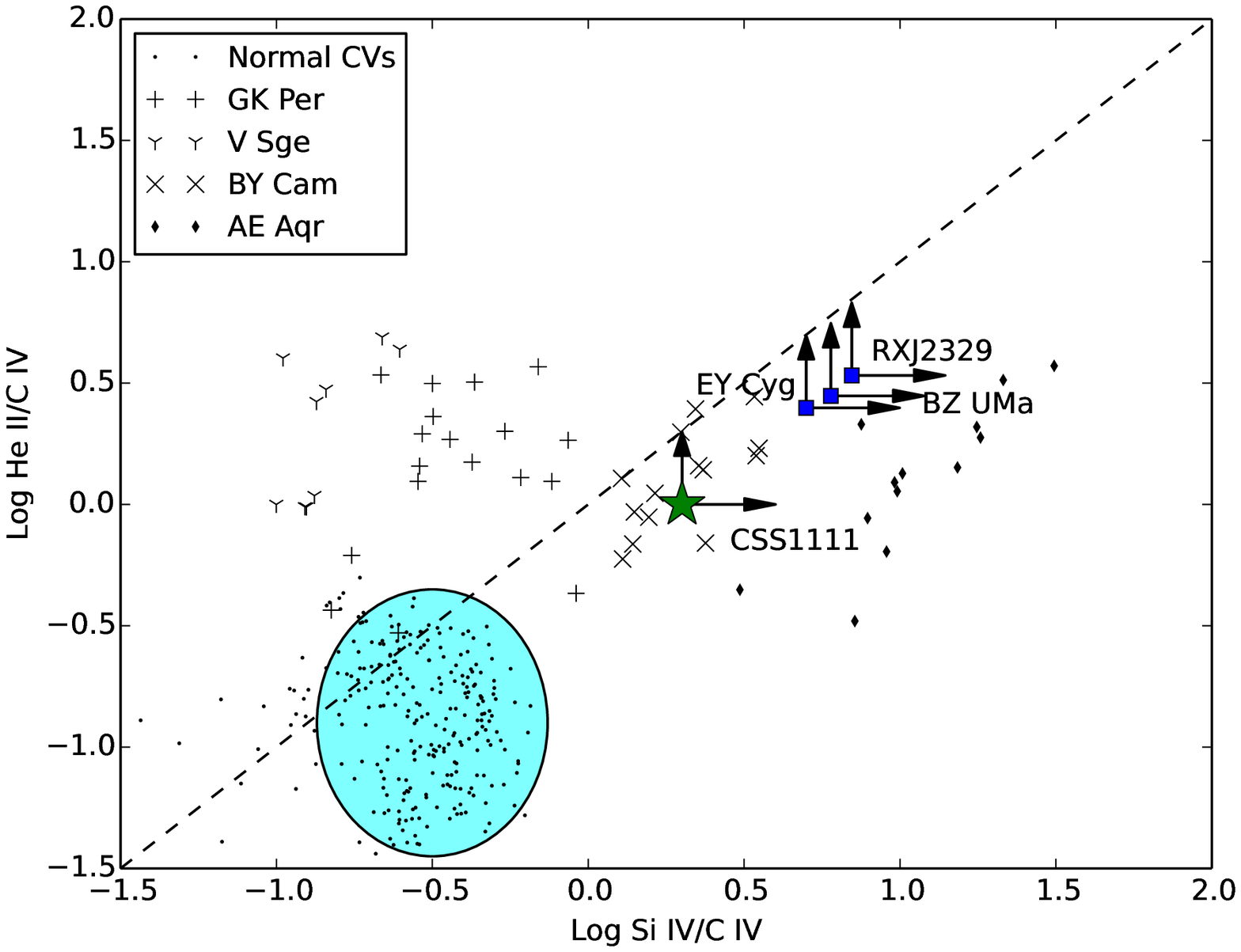}
\caption{The line ratios obtained using our COS HST data for CSS1111. It is clear that CSS1111 (green star) lies far away from the normal CV line ratio in both plots (highlighted circle), closer to the anomalous N V/ C IV inverted systems BY Cam, EY Cyg, BZ UMa, RXJ2329 (EI Psc) and AE Aqr. This suggests that the companion in this system is evolved, and the white dwarf is accreting from a hydrogen shell which has been enriched in nitrogen and depleted in carbon by the CNO process. The GK PER, V Sge, BY Cam, AE Aqr and normal CV data were kindly supplied by C. Mauche, while the data for EY Cyg, BZ Uma and RX J2329 were taken from \cite{Gansicke2003}.}\label{LineRatios}
\end{figure}

Figure \ref{LineRatios} shows our HST line ratio data combined with the line ratios of other CVs from \cite{Mauche1997} taken by the International Ultraviolet Explorer (IUE), and the lines ratios of EY Cyg, BZ UMa and EI Psc (RXJ 2329+0628) from \cite{Gansicke2003} using the Space Telescope Imaging Spectrograph (STIS) on the HST. The data show that CSS1111 lies far away from the region of normal CVs, closer to the anomalous ones identified by \cite{Gansicke2003} to within our ability to measure CIV emission from the system.

The N/C anomaly suggests that the material coming from the companion star has undergone significant nitrogen enrichment by the carbon-nitrogen process. For this material to be available for accretion, CSS1111 must have undergone TTMT to remove the outer hydrogen layers and expose the hydrogen burning layer about the core. These 2 spectral signatures, the N/C anomaly and the presence of hydrogen alongside the strong helium lines in the optical spectrum, confirm CSS1111 is an ultracompact binary produced by the EMS channel.

There are clear Si and Mg lines visible in the optical spectrum, along with the strong SiIV line visible in the HST spectrum. While the appearance of these lines is unusual, further modelling is required to quantify the abundances of these elements in CSS1111.

The SiII 6347/6371 doublet has been seen in other CVs before, most notably CP Eri \citep{Groot2001}, which is an AM CVn system with an orbital period of 29 minutes \citep{Howell1991}. Previous modelling of accretion disks suggests that, for progenitor stars with solar metallicity and optically thin He disks, the strongest metal lines in the optical should be Si II \citep{Marsh1991}. However, since CSS1111 still has substantial hydrogen visible in its spectrum when compared to AM CVn systems, the modelling done for the AM CVn systems would need to be modified to include hydrogen before being applied to this source to confirm the origin of the SiII line.

Another explanation for the appearance of these rarely seen lines could be due to the He novae that are predicted to happen in this system. For He novae where the He shell has a small mass ($<10^{-1}M_{\odot}$) and is polluted by C and N, the primary nuclear products of the detonation are $^{28}$Si and $^{40}$Ca \citep{Shen2014}. In an attempt to explain the observed N/C anomaly in the photospheres of some WDs, as opposed to in their accretion disks, \cite{Sion2014} proposed that when a nova occurs in a binary system, some of the processed material ejected from the WD ($\approx2\%-4\%$) is captured by the surface of the secondary star, and is then recycled through the accretion disk back to the surface of the WD. It is possible that a similar process is occurring in CSS1111 and other EMS channel CVs, where a He nova produces Si and other metals, which are then captured by the secondary and appear in the accretion disk.

\subsection{V418 Ser}
\begin{figure}
\epsscale{1}
\plotone{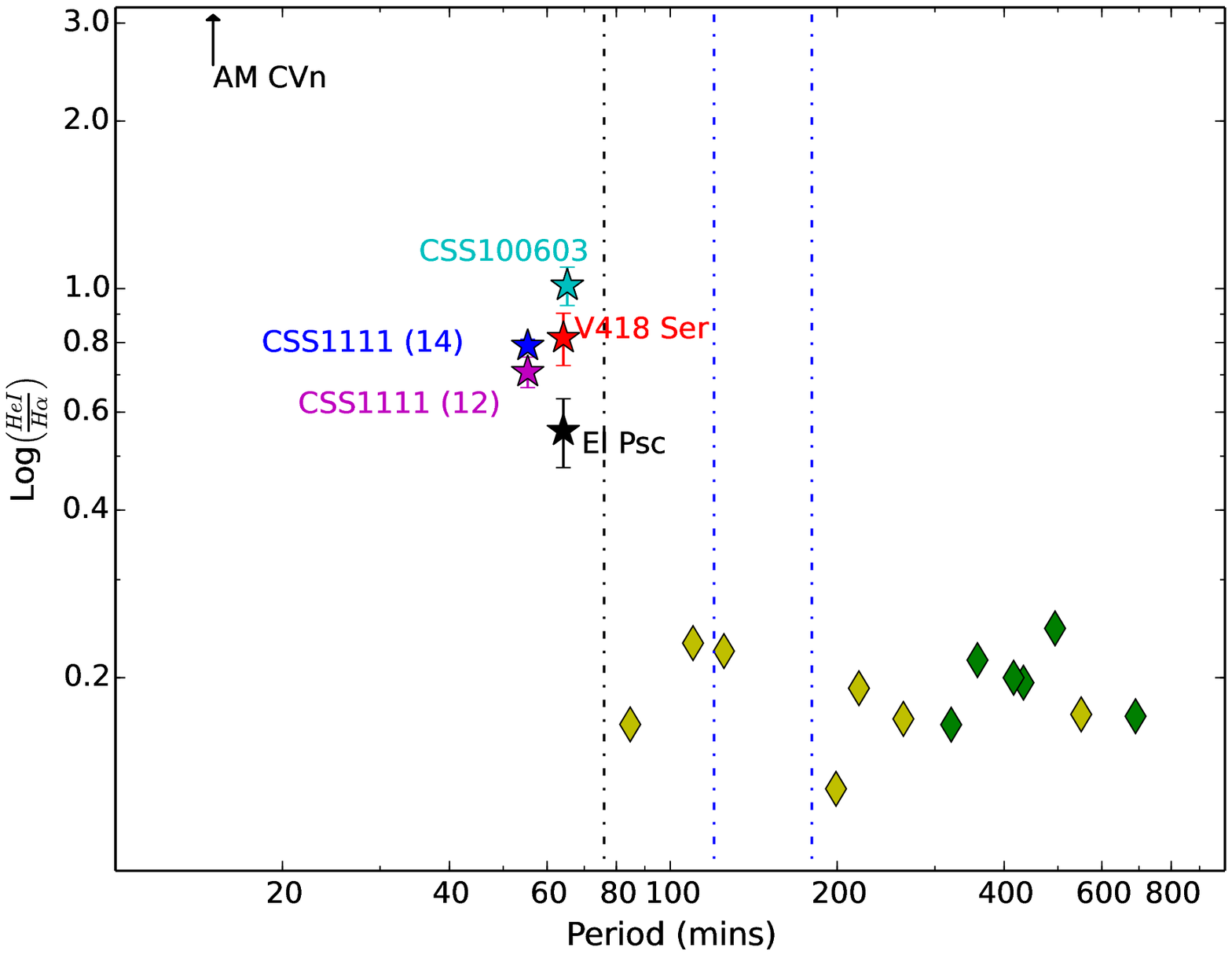}
\caption{The flux ratio of HeI 5876 over H$\alpha$ versus orbital period for different systems. CSS1111, V418 Ser and CSS 100603 have ratios between 0.7-1. We have taken the superhump period of 64.2 min to be the orbital period of V418 Ser, since the orbital period should lie close to this period. The yellow diamonds represent several long period novae and dwarf novae systems, where the ratios were taken from \cite{Williams1982} and the periods were taken from \cite{Ritter2003}. The green diamonds show the ratio for 5 CVs with periods above 5 hours, taken from \cite{Thorstensen2004}. The black dashed line shows the period minimum of 76 mins, and the 2 blue dashed lines show the period gap between 2 and 3 hours.}\label{abundanceplot}
\end{figure}

\begin{figure}
\epsscale{0.7}
\plotone{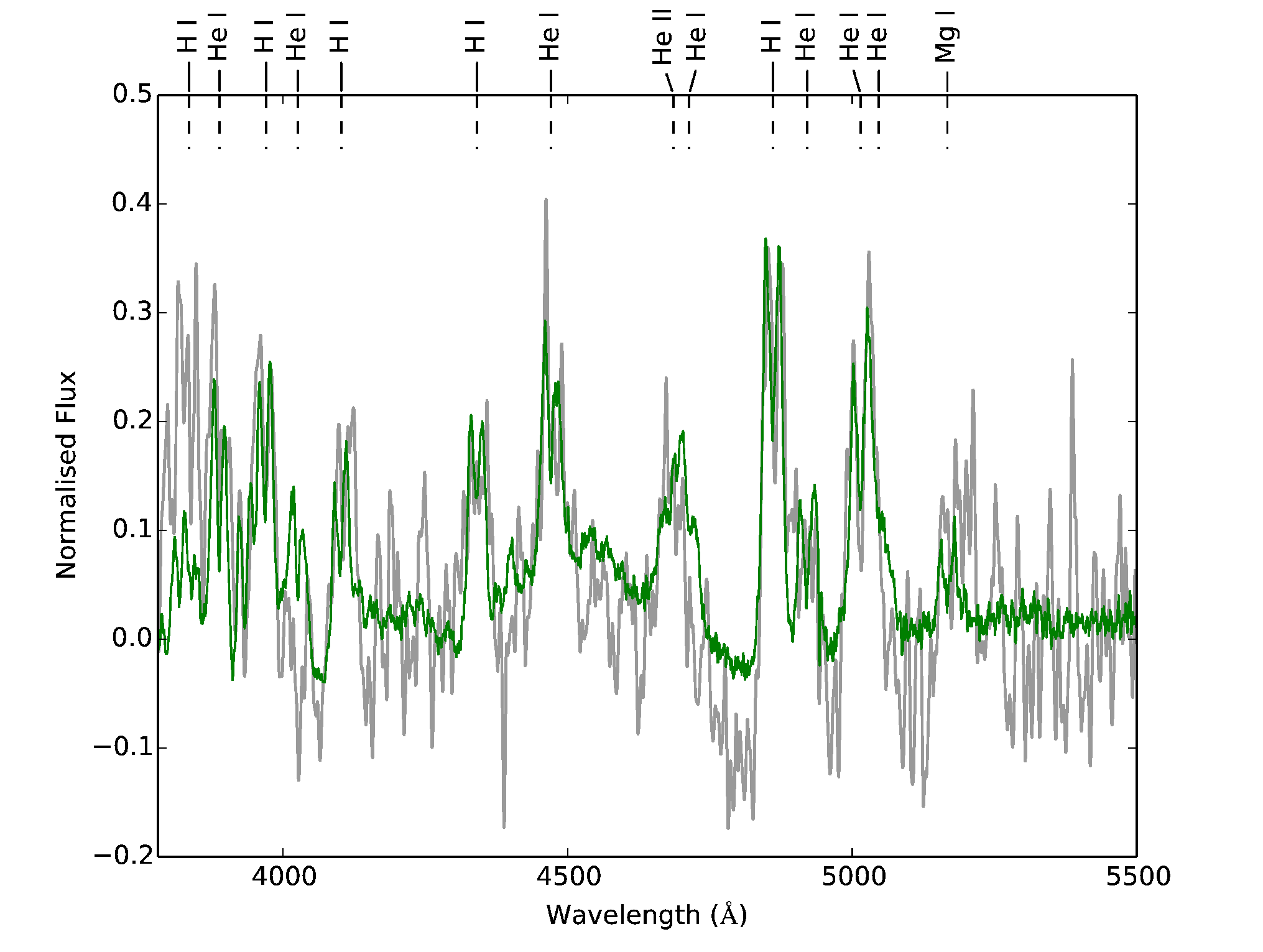}
\plotone{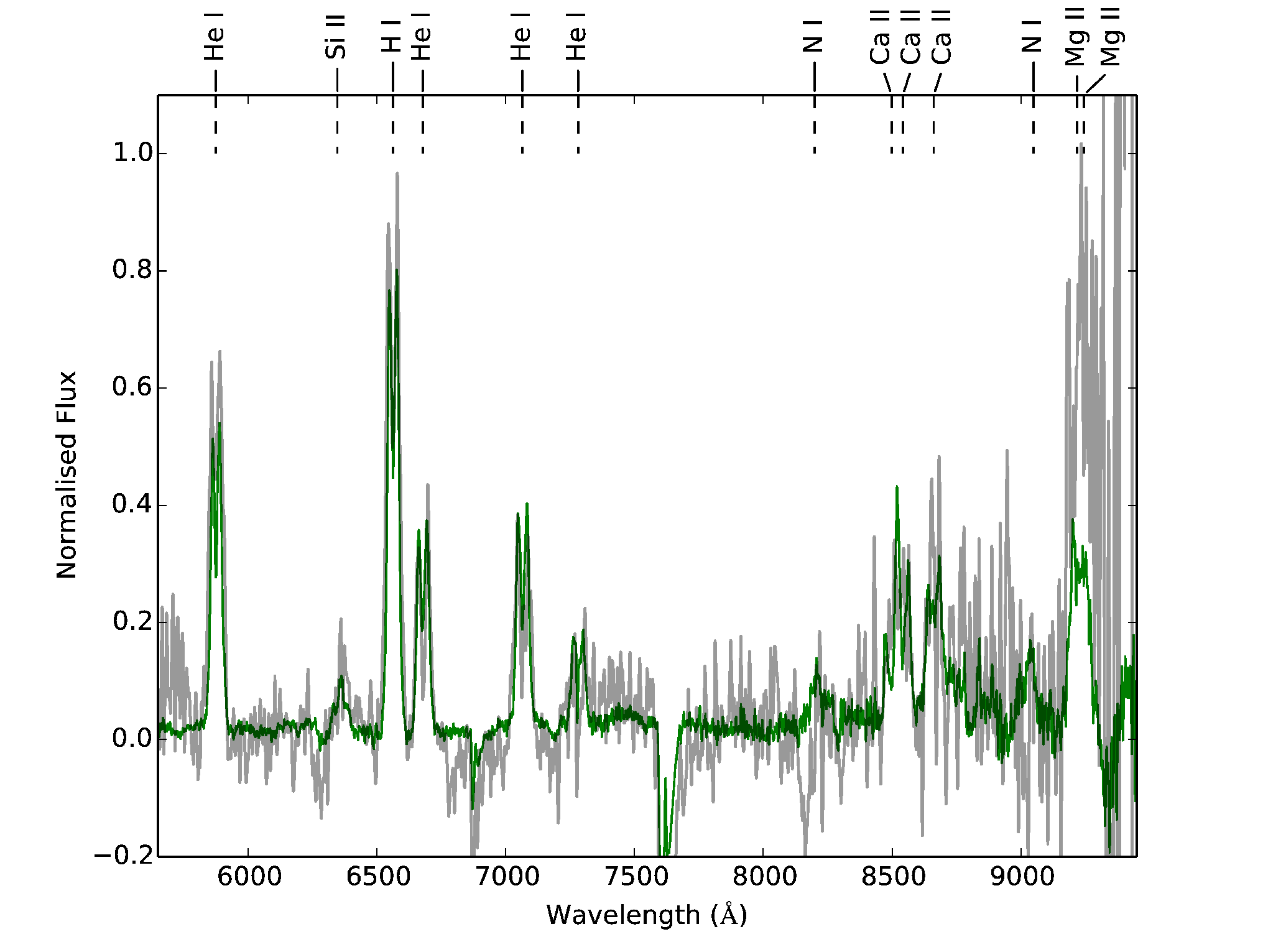}
\caption{The average spectrum obtained of V418 Ser using the LBT is shown in grey. The green line shows a scaled version of the spectrum from CSS1111 to help compare the features common to both spectra. From the similar ratios of equivalent width of HeI 5876 \AA $\:$ to H$\alpha$ and the appearance of the SiII 6347 \AA$\:$ lines, it is nearly beyond doubt that V418 Ser is a product of the evolved main sequence channel.}\label{v418_spec}
\end{figure}

We obtained 5 optical spectra of V418 Ser using the LBT on 2014 June 24 after it returned to quiescence \citep{2014ATel.6287}. The average  continuum-normalised spectrum of V418 Ser can be seen in Figure \ref{v418_spec}, alongside the spectrum of CSS1111. The main features of these spectra are nearly identical - HeI is unusually strong compared to Balmer emission, there is some HeII emission and also detectable SiII 6347. The fluxes and EW of several lines can be seen in Table \ref{opt_ew}. Figure \ref{abundanceplot} shows the flux of HeI 5876 over the flux of H$\alpha$ emission versus orbital period for CSS1111, V418 Ser and CSS 100603, another possible EMS channel CV \citep{Breedt2012}. It is clear that the HeI/H$\alpha$ ratio is very different for the EMS CVs (0.7-1.0) compared to typical CVs (0.2-0.3). Because the orbital period is shorter than the hydrogen period minimum, and the spectrum is near-identical to that of CSS1111, we classify V418 Ser as a product of the EMS channel.

\subsection{EMS Channel and Type Ia Supernovae}

Type Ia supernovae (SNIa) are thought to be detonations of carbon/oxygen (C/O) WDs as they approach the Chandrasekhar mass limit \citep{Chandrasekhar31}. Despite their use as important cosmological probes (e.g. \citealt{Riess1998}; \citealt{Perlmutter1999}), the progenitors of SNIa remain a mystery. Two mechanisms for increasing the mass of a white dwarf have long been debated without a clear resolution. Mass transfer through a ``single degenerate'' channel (SD) where the donor is a non-degenerate stellar companion \citep{Whelan1973} was the model of choice for several decades. However, slow transfer of hydrogen to the WD results in nova outbursts that appear to erode the WD mass, while the number of binaries with continuous hydrogen burning appear too low to explain the SNIa rate. Alternatively, the recent revival of the ``double degenerate'' channel, where two white dwarfs combine to reach or exceed the Chandrasekhar limit, has its own shortcomings. These include the lack of WD binaries with sufficient total mass and short orbital periods, as well as the difficulty in disrupting a WD without fusing the C/O before detonation can be achieved. (see the review by \citealt{Maoz2014}).

The SNIa progenitor problem has led to a re-examination of older models, in particular, the possibility that sub-Chandrasekhar mass WDs are the source of some SNIa. This has received support from the analysis of supernova light curves by \cite{Scalzo2014}  that suggests SNIa with fast decaying light curves eject less than 1.4 M$_\odot$ of material in their explosions. The trick is to get runaway fusion in the C/O core when the WD is not close to the Chandrasekhar mass. This can be done through a ``double detonation'' where the detonation of surface helium drives a shock toward the center that can trigger the C/O detonation (\citealt{Taam1980}; \citealt{Shen2014}). 

Early calculations required a helium shell more massive than 0.1 M$_\odot$ to achieve a detonation and the resulting explosion had a large amount of iron-group elements at high velocities so the results did not look like a SNIa (\citealt{Hoflich1996}; \citealt{Nugent1997}). Recent models suggest significantly lower mass helium shells polluted with carbon and nitrogen can detonate (\citealt{Shen2009}; \citealt{Fink2010}) and little sign of the He shell would be present in the SNIa spectra \citep{Shen2014}. \cite{Shen2014} also place an upper limit on the accretion rate in these systems of $\dot{M_{c}}<10^{-6}M_{\odot}$ yr$^{-1}$, such that stable helium burning does not occur on the surface of the WD.

If a low-mass helium shell double detonation mechanism works to make SNIa, then the products of the EMS channel are progenitors of some SNIa events. The EMS systems discussed here are transferring a larger and larger fraction of helium over time. The accretion rates in these systems are expected to be much less than $\dot{M_{c}}$, as the accretion rate for CVs with a period of $\approx55$ mins is $10^{-9}M_{\odot}$ yr$^{-1}$ \citep{Podsiadlowski2003}. The helium sinks in the WD atmosphere and survives hydrogen nova events. The over-abundance of silicon and calcium in the accretion disk may be signs that helium nova explosions have polluted the secondary, as suggested at the end of Section 6, and these eruptions are precursors to a helium detonation that could ignite the C/O in the core.

\section{Conclusions}
CSS1111 shows a strong N/C ratio, along with hydrogen and strong helium lines in the optical. These two spectral features are most likely caused by the companion in this system having just evolved off the main sequence just as mass transfer began. The main result of having an evolved main sequence star is that this system has evolved below the hydrogen period minimum, while still having hydrogen in its spectrum, unlike ultra compact systems produced by the double degenerate and helium burning channels. It is thought that as CSS1111 continues evolving, the hydrogen abundance in the system will drop as hydrogen is lost due to mass exchange. Depending on the efficiency of magnetic braking in these systems, it is possible that both CSS1111 and V418 Ser will reach a period minimum of 11 min. At this point, the hydrogen abundance in the system will be low enough that it will likely become undetectable in the optical spectrum, and the system will look like an AM CVn system formed by the double degenerate channel.

V418 Ser shows the same optical signatures as CSS1111, and has a period below the period minimum. Due to its similarity to CSS1111, we classify V418 Ser as an EMS ultra compact binary, which along with EI Psc and CSS 100603, brings the total number of CVs with strong cases for evolving by the EMS channel up to 4.

\section*{Acknowledgements}
\acknowledgments

The authors would like to thank the financial support from the Naughton Foundation, the Strategic Research Fund in University College Cork, and from the University of Notre Dame. We would also like to thank Science Foundation Ireland for financial support. We acknowledge C. Mauche for kindly supplying us with the IUE line ratio data shown in Figure \ref{LineRatios}. We would like to thank the AAVSO members for observing CSS1111 before and during HST observations, in particular Emery Erdelyi, James Foster, Franklin Guenther, Radu Gherase, Gordon Myers, Eddy Muyllaert, Velimir Popov, Roger Pickard, Richard Stanton, Daniel Taylor, Bradley Walter and Gary Walker. We would also like to thank David Lathan and Allyson Bieryla for observations made with KeplerCAM at the FLWO the night before the HST visit, and finally Henk Spruit for his Doppler mapping software. Support for program number 13427 was provided by NASA through a grant from the Space Telescope Science Institute, which is operated by the Association of Universities for Research in Astronomy, Inc., under NASA contract NAS5-26555. The CSS survey is funded by the National Aeronautics and Space Administration under Grant No. NNG05GF22G issued through the Science Mission Directorate Near-Earth Objects Observations Program. The CRTS survey is supported by the U.S.~National Science Foundation under grants AST-0909182 and AST-1313422. This paper used data obtained with the MODS spectrographs built with funding from NSF grant AST-9987045 and the NSF Telescope System Instrumentation Program (TSIP), with additional funds from the Ohio Board of Regents and the Ohio State University Office of Research. The LBT is an international collaboration among institutions in the United States, Italy and Germany. The LBT Corporation partners are: The University of Arizona on behalf of the Arizona university system; Istituto Nazionale di Astrofisica, Italy;  LBT Beteiligungsgesellschaft, Germany, representing the Max Planck Society, the Astrophysical Institute Potsdam, and Heidelberg University; The Ohio State University; The Research Corporation, on behalf of The University of Notre Dame, University of Minnesota and University of Virginia. APO is a 3.5m telescope owned and operated by the Astrophysical Research Consortium.



{\it Facilities:} \facility{LBT (MODS)}, \facility{HST (COS)}, \facility{AAVSO}, \facility{VATT}, \facility{CRTS}, \facility{APO}.

\bibliographystyle{apj.bst}
\bibliography{CSS1111_Kennedy_150907_revised}

\end{document}